\documentclass[journal]{IEEEtran}

\usepackage{rotating}
\usepackage{graphicx}
\usepackage[cmex10]{amsmath}
\usepackage{cite}
\usepackage{amssymb,amsfonts}
\usepackage{multicol}
\newtheorem{lemma}{Lemma}
\newtheorem{remark}{Remark}
\newtheorem{corollary}{Corollary}
\newtheorem{theorem}{Theorem}
\newtheorem{definition}{Definition}

\begin{document}

\title{Cumulative-Separable Codes}
\author{Sergey Bezzateev and Natalia Shekhunova
\thanks{The material in this paper was presented in part at
the  9th International Conference on Finite Fields and their
Applications,University College Dublin, July 13-17, 2009.}
\thanks{S. Bezzateev and N. Shekhunova are  with the Department
of Information Systems and Security , Saint Petersburg State
University of Airspace Instrumentation, Saint-Petersburg,
Russia, e-mail: bsv@aanet.ru, sna@delfa.net}}

\markboth{IEEE Transaction on Information Theory, Vol. , No. ,
} {Shell \MakeLowercase{\textit{et al.}}: Cumulative-Separable
Codes} \maketitle

\begin{abstract}
q-ary cumulative-separable $\Gamma(L,G^{(j)})$-codes $L=\{
\alpha \in GF(q^{m}):G(\alpha )\neq 0 \}$ and
$G^{(j)}(x)=G(x)^{j}, 1 \leq i\leq q$ are considered. The
relation between different codes from this class is
demonstrated. Improved boundaries of the minimum distance and
dimension are obtained.
\end{abstract}

\IEEEpeerreviewmaketitle
\section{Introduction}

\IEEEPARstart {A}{}t first $\Gamma(L,G)$-codes were introduced
by V.D.Goppa \cite{Goppa70} in 1970. These codes are a large
and powerful class of error correcting codes. F.J. McWilliams
and N.J. Sloane \cite{McWilliams_and_Sloan} defined these codes
as the most important class of alternant codes. It is known
that there are $\Gamma(L,G)$-codes that reach the
Gilbert-Varshamov bound and that  many $\Gamma(L,G)$-codes are
placed in the Table
 of the best known codes \cite{Br}. It is noted also that Goppa
 codes are  interesting for postquantum cryptography. There are
 four basic types of $\Gamma(L,G)$-codes: cyclic, separable,
 cumulative, and irreducible Goppa codes. In this paper we describe new type
  of $\Gamma(L,G)$-codes that we call cumulative-separable Goppa
  codes. We are motivated to study this class of Goppa codes,
  because, as it will be shown below, there are its subclasses
  that have improved estimations on minimum distance and dimension
  and that there exist codes of these subclasses that have parameters
  better than those for codes from the Table
 of the best known codes \cite{Br}.

This paper is organized as follows. In Section II we review
briefly the definitions that we will use in the paper. In
Section III we describe subclasses of cumulative-separable
$\Gamma(L,G)$-codes  with improved estimations on the dimension
and minimum distance. In Section IV the relations between codes
from different subclasses of cumulative-separable codes are
presented. In Sections V and VI theorems on estimations of the
dimension and minimum distance of considered subclasses are
presented.
\section{Class of cumulative-separable $\Gamma(L,G)$-codes}

\subsection{Cumulative Goppa codes }
\begin{definition}\cite{Goppa70}
 A Goppa code with $G(x)=(x-\alpha)^t $, where $\alpha \in GF(q^{m})$
 is called a cumulative code.
\end{definition}
It is well known\cite{Goppa70,McWilliams_and_Sloan} that the
cumulative code $\Gamma (\left\{GF(q^{m}) \setminus  \{ \alpha
\}\right\},(x-\alpha)^t)$ is equivalent to a cumulative
$\Gamma( \left\{GF(q^{m}) \setminus \{0\}\right\},x^t)$ , that,
in its tern, is equivalent to a primitive BCH-code of length
$n=q^{m} -1$ with a parity check matrix
\begin{center}
$H=\left[
\begin{array}{ccccc}
1 & \alpha^{-t} & \alpha ^{-2t} & \ldots & \alpha^{-(n-1)t} \\
1 & \alpha^{-(t-1)} & \alpha ^{-2(t-1)} & \ldots & \alpha^{-(n-1)(t-1)} \\
\ldots & \ldots & \ldots & \ldots \\
1 & \alpha^{-1} & \alpha ^{-2} & \ldots & \alpha^{-(n-1)}
\end{array}%
\right]. $
\end{center}
\begin{lemma}\label{equivalence codes power q and q-1}
 A cumulative $q$-ary  Goppa $\Gamma (L,G)$-code with  $L \subset GF(q^{m})$
 and $G(x)=x^{q}$ is equivalent to the $\Gamma (L,G^{*})$-code with  $G^{*}(x)=x^{q-1}$ .
\begin{IEEEproof}

Let us consider a parity check matrix $H^{*}$ of the $\Gamma (L,G^{*})$-code:
\begin{equation*}
H^{*}=\left[
\begin{array}{cccc}
\frac{1}{\alpha _{1}^{q-1}} & \frac{1}{\alpha _{2}^{q-1}}& \ldots
& \frac{1}{\alpha _{n}^{q-1}}
\\
\frac{\alpha _{1}}{\alpha _{1}^{q-1}} & \frac{\alpha
_{2}}{\alpha _{2}^{q-1}}&\ldots &
\frac{\alpha _{n}}{\alpha _{n}^{q-1}} \\
\ldots & \ldots & \ldots & \ldots \\
\frac{\alpha _{1}^{q-2}}{\alpha _{1}^{q-1}}& \frac{\alpha
_{2}^{q-2}}{\alpha _{2}^{q-1}} & \ldots &
\frac{\alpha_{n}^{q-2}}{\alpha _{n}^{q-1}}
\end{array}
\right].
\end{equation*}

It is clear that the row

\begin{equation*}
h_{0}=\left[
\begin{array}{cccc}
\frac{1}{\alpha _{1}^{q}} & \frac{1}{\alpha _{2}^{q}}& \ldots &
\frac{1}{\alpha _{n}^{q}}
\end{array}
\right] =  \left[
 \begin{array}{cccc}
  \frac{1}{\alpha _{1}} &
\frac{1}{\alpha _{2}}& \ldots & \frac{1}{\alpha _{n}}
\end{array}
\right]^{q}
\end{equation*}
that was obtained by raising to the $q$-th power of each
component in the last row  of the parity check matrix $H^{*}$
is also a parity check row for the $\Gamma (L,G^{*})$-code.
Thus, the matrix $H^{*}$ can be rewritten as follows:
\begin{equation*}
H^{*}=\left[
\begin{array}{cccc}
\frac{1}{\alpha _{1}^{q}} & \frac{1}{\alpha _{2}^{q}}& \ldots &
\frac{1}{\alpha _{n}^{q}}
\\
\frac{\alpha _{1}}{\alpha _{1}^{q}} & \frac{\alpha
_{2}}{\alpha_{2}^{q}}& \ldots &
\frac{\alpha _{n}}{\alpha _{n}^{q}} \\
\ldots & \ldots & \ldots & \ldots \\
\frac{\alpha _{1}^{q-1}}{\alpha _{1}^{q}}&
\frac{\alpha_{2}^{q-1}}{\alpha _{2}^{q}}& \ldots
&\frac{\alpha_{n}^{q-1}}{\alpha _{n}^{q}}
\end{array}
\right].
\end{equation*}
It is easy to see that the matrix $H^{*}$ is also a  parity check
matrix $H$ for the $\Gamma (L,G^{*})$-code.
\end{IEEEproof}
\end{lemma}

\begin{corollary}\label{equivalents_of_q-1_and_q_codes}
The cumulative $q$-ary $\Gamma (L,G)$-code with  $L \subset
GF(q^{m})$ and $G(x)=(x-\alpha)^{q}, \alpha \in GF(q^{m}) $ is
equivalent  to the  $\Gamma (L,G^{*})$-code with
$G^{*}(x)=(x-\alpha)^{q-1}$ .
\end{corollary}

\subsection{Cumulative-separable Goppa codes }
\begin{definition}
Goppa code with $G^{(j)}(x)=G(x)^{j}$ and $L \subseteq
\left\{GF(q^{m}) \setminus \{ \alpha : G(\alpha)=0 \}\right\}$,
where $G(x)$ is a separable polynomial over $GF(q^{m})$ will be
called a cumulative-separable Goppa code. Parameter $j$ is
called a cumulativity order of a cumulative-separable code.
\end{definition}

\begin{corollary}\label{cumulative_code_as_subcode}
The cumulative-separable $\Gamma (L,G^{(j)}(x))$-code can be
represented as a common subcode of  the cumulative $\Gamma
(L,G_{i}^{(j)})$-codes, where
\begin{equation*}
\begin{array}{c}
G_{i}^{(j)}(x)=(x-\beta_{i})^{j} \text{ and  }
G(x)=\prod\limits_{i=1}^{\tau}(x-\beta_{i})\;,\\
\tau= \deg G(x) ,\; \beta_{i} \in GF(q^{mt}) \; .
\end{array}
\end{equation*}
\end{corollary}

\begin{lemma}\label{code superposition redundancy}
The dimension of the cumulative-separable $\Gamma
(L,G^{(j)})$-code with $L\subseteq GF(q^{m})$ and
$G^{(j)}(x)=\prod\limits_{i=1}^{l}G_{i}^{(j)}(x)$ is determined
by the dimension of  the codes
$\Gamma_{1}(L,G_{1}^{(j)})$,$\Gamma_{2}(L,G_{2}^{(j)})$
,\ldots,$\Gamma_{l}(L,G_{l}^{(j)})$ as follows:
\begin{center}
$k \geq n-\sum\limits_{i=1}^{l} r_{i}$,
\end{center}
where $r_{i}$ is the redundancy of the
$\Gamma_{i}(L,G_{i}^{(j)})$-code.
\begin{IEEEproof}

It is clear that the parity check matrix of
$\Gamma(L,G^{(j)})$-code can be defined as:
\begin{equation*}
H=\left[
\begin{array}{c}
H_{1}\\
H_{2}\\
\ldots\\
H_{l}
\end{array}
\right],
\end{equation*}
where $ H_{i}, i=1,\ldots,l $  are parity check matrices of
$\Gamma_{i}(L,G_{i}^{(j)})$ codes.
\end{IEEEproof}
\end{lemma}

\begin{lemma}
The dimension of $q$-ary $\Gamma(L,G^{(q)})$-code of length
$n\leq q^{m}$ with $L\subseteq GF(q^{m})$ and
$G^{(q)}(x)=\left( G(x)\right)^{q}$, where $\deg G(x)=r$ and
$G(x)$ is a separable polynomial, is determined by the
inequality:
\begin{center}
$k\geq n-m(q-1)r $.
\end{center}
\begin{IEEEproof}

It follows from Lemma \ref{equivalence codes power q and q-1}
that the $\Gamma(L,G^{(q)})$-code is equivalent to the
$\Gamma(L,G^{(q-1)})$-code with
 $G^{(q-1)}(x)=\left( G(x)\right)^{q-1}$  whose
  dimension is  determined by the following inequality according to Lemma
\ref{code superposition redundancy}:
\begin{center}
$k\geq n-m(q-1)r $ .
\end{center}
\end{IEEEproof}
\end{lemma}
\section{Subclasses of cumulative-separable codes with
  improved estimations on the minimum distance and dimension }

Let us consider a family of embedded cumulative-separable
$\Gamma(L,G^{(j)})$ codes with $L\subseteq GF(q^{2l})$ and
$G^{(j)}(x)=\left( G(x)\right)^{j}$. where $j=2,..,q-1$ and the
degree of the polynomial $G(x)$ is equal to $\tau$. The parity
check matrix $H_{j}$ for $\Gamma(L,G^{(j)})$-code is written
as:
\begin{equation}
H_{j}=\left[
\begin{array}{cccc}
\frac{1}{G(\alpha _{1})^{j}} & \frac{1}{G(\alpha _{2})^{j}}&
\ldots & \frac{1}{G(\alpha _{n})^{j}}
\\
\frac{\alpha _{1}}{G(\alpha _{1})^{j}} & \frac{\alpha
_{2}}{G(\alpha _{2})^{j}}&
\ldots & \frac{\alpha _{n}}{G(\alpha _{n})^{j}} \\
\ldots & \ldots & \ldots & \ldots\\
\frac{\alpha _{1}^{\tau-1}}{G(\alpha _{1})^{j}} & \frac{\alpha
_{2}^{\tau-1}}{G(\alpha _{2})^{j}}&\ldots &
\frac{\alpha _{n}^{\tau-1}}{G(\alpha _{n})^{j}} \\
\frac{\alpha _{1}^{\tau}}{G(\alpha _{1})^{j}} & \frac{\alpha
_{2}^{\tau}}{G(\alpha _{2})^{j}}&\ldots &
 \frac{\alpha _{n}^{\tau}}{G(\alpha _{n})^{j}} \\
\frac{\alpha _{1}^{\tau+1}}{G(\alpha _{1})^{j}} & \frac{\alpha
_{2}^{\tau+1}}{G(\alpha _{2})^{j}}&\ldots &
\frac{\alpha _{n}^{\tau+1}}{G(\alpha _{n})^{j}} \\
\ldots & \ldots & \ldots  & \ldots\\
\frac{\alpha _{1}^{2\tau-1}}{G(\alpha _{1})^{j}} & \frac{\alpha
_{2}^{2\tau-1}}{G(\alpha _{2})^{j}}&\ldots &
 \frac{\alpha _{n}^{2\tau-1}}{G(\alpha _{n})^{j}} \\
\ldots & \ldots & \ldots& \ldots\\
\frac{\alpha _{1}^{(j-1)\tau-1}}{G(\alpha _{1})^{j}} &
 \frac{\alpha _{2}^{(j-1)\tau-1}}{G(\alpha _{2})^{j}}&\ldots&
 \frac{\alpha _{n}^{(j-1)\tau-1}}{G(\alpha _{n})^{j}} \\
\frac{\alpha _{1}^{(j-1)\tau}}{G(\alpha _{1})^{j}} &
\frac{\alpha _{2}^{(j-1)\tau}}{G(\alpha _{2})^{j}}&\ldots &
 \frac{\alpha _{n}^{(j-1)\tau}}{G(\alpha _{n})^{j}} \\
 \ldots & \ldots & \ldots & \ldots\\
\frac{\alpha _{1}^{j\tau-1}}{G(\alpha _{1})^{j}} & \frac{\alpha
_{2}^{j\tau-1}}{G(\alpha _{2})^{j}}&\ldots &
 \frac{\alpha _{n}^{j\tau-1}}{G(\alpha _{n})^{j}} \\
\end{array}
\right]. \label{parity-check_matrix_v1}
\end{equation}

Using the linear combination of corresponding rows of the
matrix $H_{j}$ we can present it in the following form:

\begin{equation}
H_{j}=\left[ \begin{array}{cccc}
 \frac{1}{G(\alpha _{1})^{j}}
& \frac{1}{G(\alpha _{2})^{j}}& \ldots &  \frac{1}{G(\alpha
_{n})^{j}}
\\
\frac{\alpha _{1}}{G(\alpha _{1})^{j}} & \frac{\alpha
_{2}}{G(\alpha _{2})^{j}}
&\ldots & \frac{\alpha _{n}}{G(\alpha _{n})^{j}} \\
\ldots& \ldots & \ldots  & \ldots\\
\frac{\alpha _{1}^{\tau-1}}{G(\alpha _{1})^{j}} & \frac{\alpha
_{2}^{\tau-1}}{G(\alpha _{2})^{j}}&\ldots &
\frac{\alpha _{n}^{\tau-1}}{G(\alpha _{n})^{j}} \\
\frac{1}{G(\alpha _{1})^{j-1}} & \frac{1}{G(\alpha
_{2})^{j-1}}&
\ldots & \frac{1}{G(\alpha _{n})^{j-1}} \\
\frac{\alpha _{1}}{G(\alpha _{1})^{j-1}} & \frac{\alpha
_{2}}{G(\alpha _{2})^{j-1}}&\ldots &
\frac{\alpha _{n}}{G(\alpha _{n})^{j-1}} \\
\ldots & \ldots & \ldots & \ldots\\
\frac{\alpha _{1}^{\tau-1}}{G(\alpha _{1})^{j-1}} &
 \frac{\alpha _{2}^{\tau-1}}{G(\alpha _{2})^{j-1}}&\ldots&
 \frac{\alpha _{n}^{\tau-1}}{G(\alpha _{n})^{j-1}} \\
 \ldots & \ldots & \ldots&\ldots\\
\frac{\alpha _{1}^{\tau-1}}{G(\alpha _{1})^{2}} & \frac{\alpha
_{2}^{\tau-1}}{G(\alpha _{2})^{2}}&\ldots &
 \frac{\alpha _{n}^{\tau-1}}{G(\alpha _{n})^{2}} \\
\frac{1}{G(\alpha _{1})} & \frac{1}{G(\alpha _{2})}&\ldots &
 \frac{1}{G(\alpha _{n})} \\
 \ldots& \ldots & \ldots &  \ldots\\
\frac{\alpha _{1}^{\tau-1}}{G(\alpha _{1})} &
\frac{\alpha _{2}^{\tau-1}}{G(\alpha _{2})}&\ldots &
 \frac{\alpha _{n}^{\tau-1}}{G(\alpha _{n})} \\
\end{array}
\right]. \label{parity-check_matrix_v2}
\end{equation}

Thus, the parity check matrix of the $\Gamma(L,G^{(j+1)})$-code
is obtained from that of the preceding $\Gamma(L,G^{(j)})$-code
by writing up a submatrix $h_{j+1}$:
\begin{equation*}
h_{j+1}=\left[
 \begin{array}{cccc}
\frac{1}{G(\alpha _{1})^{j+1}} & \frac{1}{G(\alpha _{2})^{j+1}}&\ldots&
 \frac{1}{G(\alpha _{n})^{j+1}} \\
\frac{\alpha _{1}}{G(\alpha _{1})^{j+1}} &
 \frac{\alpha _{2}}{G(\alpha _{2})^{j+1}}&\ldots &
   \frac{\alpha _{n}}{G(\alpha _{n})^{j+1}} \\
 \ldots& \ldots & \ldots& \ldots\\
\frac{\alpha _{1}^{\tau-1}}{G(\alpha _{1})^{j+1}} &
\frac{\alpha _{2}^{\tau-1}}{G(\alpha _{2})^{j+1}}&\ldots &
\frac{\alpha _{n}^{\tau-1}}{G(\alpha _{n})^{j+1}} \\
\end{array}
\right].
\end{equation*}
Respectively, the parity check matrix $H_{j}$ can be rewritten
as follows:
\begin{center}
$H_{j}=\left[ \begin{array}{c}
h_{1}\\
h_{2}\\
\ldots\\
h_{j}\\
\end{array}
\right].$
\end{center}
We can obtain the following corollary from the above relations.
\begin{corollary} ~\label{subcodes_in_comul_separable_codes}
The cumulative-separable  $\Gamma(L,G^{(j)})$-code is a subcode
of the cumulative-separable $\Gamma(L,G^{(j-1)})$-code.
\end{corollary}

Consider now cumulative-separable codes with the improved
estimation on dimension that are obtained from the following
separable Goppa codes \cite{Bezz7,Bezz10}:
 $\begin{array}{rl}
        \Gamma_{1}=\Gamma (L_{1},G_{1}), &
        L_{1} = \{GF(t^{2}) \backslash \{\alpha :G_{1}(\alpha)=0 \}\} \\
         \text{and} &G_{1}(x)=x^{t-1}+1;
       \end{array}$
 $\begin{array}{rl}
         \Gamma_{1}^{\ast }=\Gamma (L_{1}^{\ast },G_{1}), &
         L_{1}^{\ast } = \{L_{1} \backslash \{0\} \} \\
         \text{and} &  G_{1}(x)=x^{t-1}+1;
       \end{array}$
$\begin{array}{rl}
         \Gamma_{2}=\Gamma(L_{2},G_{2}) ,&
         L_{2} = L_{1}^{\ast }\\
         \text{and} & G_{2}(x)=x^{t}+x;
       \end{array}$
 $\begin{array}{rl}
        \Gamma_{3}=\Gamma(L_{3},G_{3}),  & L_{3} =
    \{GF(t^{2})\backslash \{\alpha :G_{3}(\alpha)=0 \}\} \\
         \text{and} &G_{3}(x)=x^{t}+x+1;
       \end{array}$
 $\begin{array}{rl}
         \Gamma_{4}^{\ast }=\Gamma(L_{4}^{\ast },G_{4}), &
         L_{4}^{\ast } = GF(t^{2})\setminus \left\{ \{\alpha :G_{4
}(\alpha)=0 \} \bigcup\{0\} \right\} \\
        \text{and} & G_{4}(x)=x^{t}+x^{t-1}+1;
       \end{array}$
 $\begin{array}{rl}
         \Gamma_{5}=\Gamma(L_{5},G_{5}), & L_{5} = L_{4}^{\ast } \\
         \text{and}& G_{5}(x)=x^{t+1}+x^{t}+x;
       \end{array}$
$\begin{array}{rl}
         \Gamma_{6}=\Gamma(L_{6},G_{6}), & L_{6} =
    \{GF(t^{2})\setminus \{\alpha :G_{6}(\alpha)=0 \}\} \\
         \text{and} & G_{6}(x)=x^{t+1}+1.
       \end{array}$

As it has been shown earlier in \cite{Bezz7,Bezz10}, these
codes are related by following relations that allow us to speak
about a chain of equivalent and embedded codes.
\begin{itemize}
  \item  The $\Gamma _{1}^{*}$-code is obtained from the $\Gamma _{1}$-code
  by shortening in information symbol corresponding to the
  numerator  $\left\{0\right\}$ from the set $L_{1}$.
  The parameters of these codes are related by following
  relations: $n_{1}^{*} = n_{1}-1 = (t^{2}-t+1)-1$,
$r_{1}^{*}=r_{1}$, $k_{1}^{*}= k_{1}-1$, $d_{1}^{*}\geq
d_{1}=t$.
  \item The code $\Gamma _{2}$ is a subcode of $\Gamma _{1}^{*}$.
   $L_{2}=L_{1}^{*}=L_{1}\setminus\left\{0\right\}$, i.e.,
 $n_{2}=n_{1}^{*}=n_{1}-1$ and
   $G_{2}(x) = xG_{1}(x)$.
The parity check matrices of these codes are related by the following relation:
$$
H_{2}= \left[
\begin{array}{c}
\frac{1}{G_{2}(\alpha _{i_{1}})}  \ldots \frac{1}{G_{2}(\alpha
_{i_{n_{2}}})} \\
H_{1}^{*}
\end{array}%
\right],
$$
where $H_{1}^{*}$ is the parity check matrix of the  code
$\Gamma_{1}^{*}$, with the redundancy $r_{1}^{*}=r_{1}$. Hence,
the redundancy of the code $\Gamma _{2}$ is determined by the
inequality $r_{2}\leq r_{1} + l.$ The minimum distance is $d_{2}
\geq deg G_{2}(x) +1 =t+1$.
  \item The code $\Gamma_{3}$ is equivalent to code $\Gamma _{2}$
  up to the permutation
\begin{center}
$ \gamma\rightarrow \gamma+\beta,\text{ where the element }
\beta \in GF(t^{2}) \text{ is such that } \beta^{t}+\beta=1
$ and $\gamma \in L_{2}$.
\end{center}
It follows that $n_{3}=n_{2},\; k_{3}=k_{2},\;
d_{3}=d_{2}$.
\item
 An auxiliary code $C_{3}^{*}$ is obtained by shortening
    of the code  $\Gamma_{3}$ in a redundancy symbol corresponding to  the numerator
    $\{0\} \in L_{3}$.
Thus, the parity check matrix  $H^{*}_{3}$ of the code
$C_{3}^{*}$ results from deleting the first
    unit row and the last column of the  parity check  matrix $H_{3}$
     of the code $\Gamma_{3}$:
$$
H_{3}= \left[
\begin{array}{cccc}
1& \ldots&1&1\\
 &H_{3}^{*}& &0\\
 \end{array}
 \right] .
 $$
 Therefore, $L_{3}^{*}=L_{3} \setminus \{0\}$,\;
 $n_{3}^{*}=n_{3}-1,\; k_{3}^{*}=k_{3}, \; d_{3}^{*} \leq
 d_{3}$.
 \item  The auxiliary code $C_{3}^{*}$ is equivalent to the code
  $\Gamma_{4}^{*}$ up to permutation
  $ \beta \rightarrow \frac{1}{\beta}, \text{ where element } \beta
\in L_{3}^{*} $ and, hence, $n_{4}^{*}=n_{3}^{*}, \;
k_{4}^{*}=k_{3}^{*},\; d_{4}^{*}=d_{3}^{*}$.
\item The code $\Gamma _{5}$ is a subcode of the  code
    $\Gamma _{4}^{*}$  and their parity check matrices are
    related as follows:
\begin{center}
$ H_{5}=\left[
\begin{array}{ccc}
\frac{1}{G_{5}(\alpha _{i_{1}})} & \ldots & \frac{1}{G_{5}(\alpha
_{i_{n_{5}}})}  \\
 & H_{4}^{*}  & \\
\end{array}
\right].
 $
\end{center}
The code $\Gamma_{5}$ has the following parameters:
$n_{5}=n_{4}^{*}, \; k_{5}=k_{4}^{*}-l, \; d_{5}\geq
d_{4}^{*}$.
\item The codes $\Gamma _{5}$ and $\Gamma _{6}$ are equivalent
    up to the permutation $\beta \rightarrow \gamma\beta-1$,
    where    $\gamma \in GF(t^{2})$ and $\gamma^{t+1}+1=0$ , $\beta \in L_{5}$.
     Thus, $n_{6}=n_{5}$ , \; $k_{6}=k_{5}$ and $d_{5}=d_{6}=t+2$.
\end{itemize}
Let us transfer now the relations between the above subclasses
of separable codes to the case of cumulative-separable codes
obtained from these codes.

\section{Relations between cumulative-separable codes from different subclasses}\label{section_chain_comulative-separable codes}
\subsection{ Relation between the cumulative-separable codes
$\Gamma_{1}^{*(i)}$ and $\Gamma_{1}^{(i)}$ }
\begin{itemize}
\item $\Gamma_{1}^{(i)}=\Gamma (L_{1},G_{1}^{(i)})$ with $L_{1}
    = \left\{GF(t^{2})\backslash\{\alpha :G_{1}(\alpha)=0\}\right\}$, $n_{1}^{(i)}=n_{1}$ and
     $G_{1}^{(i)}(x)=(x^{t-1}+1)^{i}$ ;
\item $\Gamma_{1}^{\ast(i) }=\Gamma (L_{1}^{\ast
    }, \; G_{1}^{(i)})$  with    $L_{1}^{\ast } = \{L_{1} \backslash \{0\} \}
     ,\;  n_{1}^{\ast(i)}=n_{1}^{\ast}$ and
    $G_{1}^{(i)}(x)=(x^{t-1}+1)^{i}$ ;
\end{itemize}
  It follows from the above considered definition of these codes that the
  $\Gamma_{1}^{*(i)}$-code is obtained from the $\Gamma_{1}^{(i)}$-code
   by shortening in an information symbol corresponding to the numerator $\{0\}$.

  Parameters of these codes are related as:
  $n_{1}^{*}=n_{1}-1=t^{2}-t, \;
  r_{1}^{*(i)}=r_{1}^{(i)}, \; k_{1}^{*(i)}=k_{1}^{(i)}-1, \; d_{1}^{*(i)}\geq
  d_{1}^{(i)} $.
\subsection{Relation between the cumulative-separable codes
  $\Gamma_{2}^{(i)}$ and  $\Gamma_{1}^{*(i)}$}~\label{subsection_chain_comulative_separablecodes}
$$\begin{array}{rcl}
\Gamma_{2}^{(i)}=\Gamma(L_{2},G_{2}^{(i)})&
\text{ with } &L_{2} = L_{1}^{\ast } , \\
n_{2}^{(i)}=n_{2} &\text{ and  }& G_{2}^{(i)}(x)=(x^{t}+x)^{i}
.
\end{array}$$
Two cases should be considered here. They are determined by
the cumulativity order $i$:

$\boldsymbol{i <q-1 :}$  In this
    case the
    $\Gamma
    _{2}^{(i)}$-code is  a subcode of the $\Gamma
    _{1}^{*(i)}$-code.
   $L_{2}=L_{1}^{*}=L_{1}\setminus\left\{0\right\}$, i.e. ,
   $n_{2}=n^{*}_{1}=n_{1}-1$ and  $G_{2}^{(i)}(x) = x^{i}G_{1}^{(i)}(x)$.
\begin{lemma}~\label{lemma_i<q-1_G1*_G2}
   The parity check matrices of  the  codes  $\Gamma_{2}^{(i)}$
   and    $\Gamma_{1}^{*(i)}$ are related as follows:
$$
H_{2}^{(i)}= \left[
\begin{array}{ccc}
\frac{1}{G_{2}^{(i)}(\alpha _{i_{1}})} & \ldots&
\frac{1}{G_{2}^{(i)}(\alpha
_{i_{n_{2}}})} \\
& H_{1}^{*(i)} &
\end{array}
\right],
$$
where $H_{1}^{*(i)}$ is the parity check matrix of the
$\Gamma_{1}^{*(i)}$-code.

\begin{IEEEproof}

We use representation
(\ref{parity-check_matrix_v1}) of the parity check matrix of the cumulative-separable code $\Gamma_{2}^{(i)}$:
\begin{equation}
\begin{array}{r}
  H_{2}^{(i)}  = \left[
\begin{array}{ccc}
\frac{1}{\alpha _{1}^{i}(\alpha _{1}^{t-1}+1)^{i}} & \ldots &
\frac{1}{\alpha _{n_{2}}^{i}(\alpha _{n_{2}}^{t-1}+1)^{i}}
\\
\frac{1}{\alpha _{1}^{i-1}(\alpha _{1}^{t-1}+1)^{i}} &
 \ldots &
\frac{1}{\alpha _{n_{2}}^{i-1}(\alpha _{n_{2}}^{t-1}+1)^{i}}
\\
\ldots &  \ldots & \ldots \\
\frac{1}{\alpha _{1}(\alpha _{1}^{t-1}+1)^{i}} & \ldots &
\frac{1}{\alpha _{n_{2}}(\alpha _{n_{2}}^{t-1}+1)^{i}}
\\
\frac{1}{(\alpha _{1}^{t-1}+1)^{i}} &  \ldots & \frac{1}{(\alpha
_{n_{2}}^{t-1}+1)^{i}}
\\
\ldots & \ldots  & \ldots \\
\frac{\alpha _{1}^{i(t-1)-1}}{(\alpha _{1}^{t-1}+1)^{i}} & \ldots
& \frac{\alpha _{n_{2}}^{i(t-1)-1}}{(\alpha
_{n_{2}}^{t-1}+1)^{i}}
\end{array}
\right] \\
 =  \left[
\begin{array}{ccc}
\frac{1}{\alpha _{1}^{i}(\alpha _{1}^{t-1}+1)^{i}} &
 \ldots & \frac{1}{\alpha _{n_{2}}^{i}(\alpha _{n_{2}}^{t-1}+1)^{i}}
\\
\frac{1}{\alpha _{1}^{i-1}(\alpha _{1}^{t-1}+1)^{i}} &
 \ldots & \frac{1}{\alpha _{n_{2}}^{i-1}(\alpha _{n_{2}}^{t-1}+1)^{i}}
\\
\ldots & \ldots  & \ldots \\
\frac{1}{\alpha _{1}(\alpha _{1}^{t-1}+1)^{i}} &
 \ldots & \frac{1}{\alpha _{n_{2}}(\alpha _{n_{2}}^{t-1}+1)^{i}}
\\
 & H_{1}^{*(i)} &
 \end{array} \right].
\end{array}
  \label{parity-check_matrix_comul_G2}
\end{equation}
It is easy to see that for component-wise raising to a power
$t$ of any row
$$\left[
\begin{array}{ccc}\frac{1}{\alpha _{1}^{j}(\alpha
_{1}^{t-1}+1)^{i}} &  \ldots & \frac{1}{\alpha _{n_{2}}^{j}(\alpha
_{n_{2}}^{t-1}+1)^{i}} \end{array} \right] , 0<j<i $$
 of the matrix $H_{2}^{(i)}$  the corresponding row of the submatrix
$H_{1}^{*(i)}$ is obtained:
$$
\begin{array}{r}
\left[
\begin{array}{ccc}\frac{1}{\alpha _{1}^{j}(\alpha
_{1}^{t-1}+1)^{i}} &  \ldots & \frac{1}{\alpha _{n_{2}}^{j}(\alpha
_{n_{2}}^{t-1}+1)^{i}} \end{array} \right]^{t}=\\
\left[
\begin{array}{ccc}\frac{1}{\alpha _{1}^{jt}(\alpha
_{1}^{1-t}+1)^{i}} &  \ldots & \frac{1}{\alpha
_{n_{2}}^{jt}(\alpha _{n_{2}}^{1-t}+1)^{i}} \end{array}
\right]=\\
\left[
\begin{array}{ccc}\frac{\alpha _{1}^{t(i-j)-i}}{(\alpha
_{1}^{t-1}+1)^{i}} &  \ldots & \frac{\alpha
_{n_{2}}^{t(i-j)-i}}{(\alpha _{n_{2}}^{t-1}+1)^{i}} \end{array}
\right].
\end{array}
$$
Thus, the parity check matrix of the $\Gamma_{2}^{(i)}$-code
can be written as
\begin{equation}
H_{2}^{(i)}=\left[
\begin{array}{ccc}\frac{1}{\alpha _{1}^{i}(\alpha _{1}^{t-1}+1)^{i}} &
 \ldots & \frac{1}{\alpha _{n_{2}}^{i}(\alpha _{n_{2}}^{t-1}+1)^{i}}
\\
 & H_{1}^{*(i)} &
 \end{array} \right] ,\label{parity-check_matrix_comul_G2_G1*}
\end{equation}
where the first row consists of the elements from $GF(t)$ only:
$$
\begin{array}{r}
\left[
\begin{array}{ccc}\frac{1}{\alpha _{1}^{i}(\alpha _{1}^{t-1}+1)^{i}} &
 \ldots & \frac{1}{\alpha _{n_{2}}^{i}(\alpha _{n_{2}}^{t-1}+1)^{i}}\end{array}
 \right]=\\
 \left[
\begin{array}{ccc}\frac{1}{(\alpha _{1}^{t}+\alpha _{1})^{i}} &
 \ldots & \frac{1}{(\alpha _{n_{2}}^{t}+\alpha _{n_{2}})^{i}}\end{array}
 \right] = \\
 \left[
\begin{array}{ccc}
\left(\frac{1}{G_{2}(\alpha _{i_{1}})}\right)^{i} & \ldots&
\left(\frac{1}{G_{2}(\alpha _{i_{n_{2}}})}\right)^{i}
\end{array}
\right].
\end{array}
$$
Hence, the redundancy of the  $\Gamma_{2}^{(i)}$-code is
determined by the inequality $r_{2}^{(i)}\leq r_{1}^{(i)} + l$
.
\end{IEEEproof}
\end{lemma}

$\boldsymbol{i = q-1 , i = q\; :}$  By using Corollary
\ref{equivalents_of_q-1_and_q_codes}
  and Corollary \ref{cumulative_code_as_subcode} we obtain that the
    $\Gamma
    _{2}^{(q-1)}$-code is equivalent to the $\Gamma
    _{2}^{(q)}$-code and the $\Gamma _{1}^{*(q-1)}$-code is equivalent to the
$\Gamma _{1}^{*(q)}$-code.
    Let us prove now the equivalency of the codes $\Gamma_{2}^{(q-1)}$ and $\Gamma _{1}^{*(q-1)}$.

\begin{lemma}
   Parity check matrices of the codes $\Gamma_{2}^{(q-1)}$ and
   $\Gamma_{1}^{*(q-1)}$ are equal:
$$
H_{2}^{(q-1)}=  H_{1}^{*(q-1)}\; .
$$
\begin{IEEEproof}

Using Lemma \ref{lemma_i<q-1_G1*_G2}, the relation between the
parity check matrices of the  codes $\Gamma_{2}^{(q-1)}$ and
$\Gamma_{1}^{*(q-1)}$ can be written as:
\begin{equation*}
H_{2}^{(q-1)}=\left[
\begin{array}{ccc}
\frac{1}{\alpha _{1}^{q-1}(\alpha _{1}^{t-1}+1)^{q-1}} &
 \ldots & \frac{1}{\alpha _{n_{2}}^{q-1}(\alpha _{n_{2}}^{t-1}+1)^{q-1}}
\\
 & H_{1}^{*(q-1)} &
 \end{array} \right].
\end{equation*}
 As the codes  $\Gamma _{1}^{*(q-1)}$ and  $\Gamma _{1}^{*(q)}$
are equivalent, then their parity check matrices are equal:
$H_{1}^{*(q-1)}= H_{1}^{*(q)} $. We show that the row
 $$\left[
\begin{array}{ccc}
\frac{1}{\alpha _{1}^{q-1}(\alpha _{1}^{t-1}+1)^{q-1}} &
 \ldots & \frac{1}{\alpha _{n_{2}}^{q-1}(\alpha _{n_{2}}^{t-1}+1)^{q-1}}
 \end{array} \right]$$
 of the matrix  $H_{2}^{(q-1)}$ can be obtained as the linear
 combination of the row of the matrix  $H_{1}^{*(q)}$ :
 $$\left[\begin{array}{ccc}
\frac{\alpha _{1}^{t-q}}{(\alpha _{1}^{t-1}+1)^{q}} &
 \ldots & \frac{\alpha _{n_{2}}^{t-q}}{(\alpha _{n_{2}}^{t-1}+1)^{q}}\end{array}
 \right]$$
 and its $t$-th power:
$$
\begin{array}{r}
\left(\left[
\begin{array}{ccc}
\frac{\alpha _{1}^{t-q}}{(\alpha _{1}^{t-1}+1)^{q}} &
 \ldots & \frac{\alpha _{n_{2}}^{t-q}}{(\alpha _{n_{2}}^{t-1}+1)^{q}}
 \end{array} \right]\right)^{t} + \\
 \left[
\begin{array}{ccc}
\frac{\alpha _{1}^{t-q}}{(\alpha _{1}^{t-1}+1)^{q}} &
 \ldots & \frac{\alpha _{n_{2}}^{t-q}}{(\alpha _{n_{2}}^{t-1}+1)^{q}}
 \end{array} \right]= \\
 \left[
\begin{array}{ccc}
\frac{1}{\alpha _{1}^{q-1}(\alpha _{1}^{t-1}+1)^{q-1}} &
 \ldots & \frac{1}{\alpha _{n_{2}}^{q-1}(\alpha _{n_{2}}^{t-1}+1)^{q-1}}\end{array}
 \right].
\end{array}
$$
Indeed,
$$
\begin{array}{r}
\left(\left[
\begin{array}{ccc}
\frac{\alpha _{1}^{t-q}}{(\alpha _{1}^{t-1}+1)^{q}} &
 \ldots & \frac{\alpha _{n_{2}}^{t-q}}{(\alpha _{n_{2}}^{t-1}+1)^{q}}\end{array}
 \right]\right)^{t}= \\
 \left[
\begin{array}{ccc}
\frac{\alpha _{1}^{1-qt}}{(\alpha _{1}^{1-t}+1)^{q}} &
 \ldots & \frac{\alpha _{n_{2}}^{1-qt}}{(\alpha _{n_{2}}^{1-t}+1)^{q}}\end{array}
 \right]=\\
 \left[
\begin{array}{ccc}
\frac{1}{\alpha _{1}^{q-1}(\alpha _{1}^{t-1}+1)^{q}} &
 \ldots & \frac{1}{\alpha _{n_{2}}^{q-1}(\alpha _{n_{2}}^{t-1}+1)^{q}}\end{array}
 \right]
\end{array}
$$
and
$$
\begin{array}{r}
\left[
\begin{array}{ccc}
\frac{1}{\alpha _{1}^{q-1}(\alpha _{1}^{t-1}+1)^{q}} &
 \ldots & \frac{1}{\alpha _{n_{2}}^{q-1}(\alpha _{n_{2}}^{t-1}+1)^{q}}\end{array}
 \right] +\\
\left[
\begin{array}{ccc}
\frac{\alpha _{1}^{t-q}}{(\alpha _{1}^{t-1}+1)^{q}} &
 \ldots & \frac{\alpha _{n_{2}}^{t-q}}{(\alpha _{n_{2}}^{t-1}+1)^{q}}\end{array}
 \right]= \\
 \left[
\begin{array}{ccc}
\frac{\alpha _{1}^{t-1}+1}{\alpha _{1}^{q-1}(\alpha
_{1}^{t-1}+1)^{q}} &
 \ldots & \frac{\alpha _{n_{2}}^{t-1}+1}{\alpha _{n_{2}}^{q-1}(\alpha _{n_{2}}^{t-1}+1)^{q}}\end{array}
 \right] =\\
 \left[
\begin{array}{ccc}
\frac{1}{\alpha _{1}^{q-1}(\alpha _{1}^{t-1}+1)^{q-1}} &
 \ldots & \frac{1}{\alpha _{n_{2}}^{q-1}(\alpha _{n_{2}}^{t-1}+1)^{q-1}}\end{array}
 \right].
  \end{array}
$$.

\end{IEEEproof}
\end{lemma}
Hence, the redundancy of the $\Gamma_{2}^{(q-1)}$-code
coincides with that of the $\Gamma_{1}^{*(q-1)}$-code:
$$r_{2}^{(q-1)}= r_{1}^{*(q-1)} .$$
\subsubsection{Relation between the cumulative-separable codes
 $\Gamma_{2}^{(i)}$ and  $\Gamma_{3}^{(i)}$}

Let us define:
  $$
  \begin{array}{l}
  \Gamma_{3}^{(i)}=\Gamma(L_{3},G_{3}^{(i)}) \text{ with } L_{3} =
\{GF(t^{3}) \backslash \{\alpha : G_{3}(\alpha)=0\}\} ,\\
n_{3}^{(i)}=n_{3} \text{ and } G_{3}^{(i)}(x)=(x^{t}+x+1)^{i} .
 \end{array}
$$
It follows that the codes $\Gamma_{3}^{(i)}$ and $\Gamma _{2}^{(i)}$ are equivalent for all $i\geq 1$ with up to permutation
$$
\begin{array}{c}
 \gamma\rightarrow \gamma+\beta,\text{ where element } \beta
\in GF(t^{2}) \\
\text{ is such that } \beta^{t}+\beta=1  $ and $\gamma \in
L_{2}.
\end{array} $$
Hence, $n_{3}=n_{2},\; k_{3}^{(i)}=k_{2}^{(i)},\;
d_{3}^{(i)}=d_{2}^{(i)}$.

\subsection{Relation between the cumulative-separable
  $\Gamma_{3}^{(i)}$-code and auxiliary $ C_{3}^{*(i)}$-code}

Let us use representation (\ref{parity-check_matrix_v1}) of the
parity check matrix of the $\Gamma_{3}^{(i)}$-code:
\begin{equation}
H_{3}^{(i)}=\left[
\begin{array}{cccc}
\frac{1}{(\alpha _{1}^{t}+\alpha_{1}+1)^{i}} & \ldots &
\frac{1}{(\alpha _{n_{3}-1}^{t}+\alpha _{n_{3}-1}+1)^{i}}& 1
\\
\frac{\alpha _{1}}{(\alpha _{1}^{t}+\alpha_{1}+1)^{i}} &
 \ldots &
\frac{\alpha _{n_{3}-1}}{(\alpha _{n_{3}-1}^{t}+\alpha
_{n_{3}-1}+1)^{i}}& 0
\\
\ldots &  \ldots & \ldots &0\\
\frac{\alpha _{1}^{i}}{(\alpha _{1}^{t}+\alpha_{1}+1)^{i}} &
 \ldots &
\frac{\alpha _{n_{3}-1}^{i}}{(\alpha _{n_{3}-1}^{t}+\alpha
_{n_{3}-1}+1)^{i}}& 0\\
\ldots &  \ldots & \ldots &0\\
 \frac{\alpha _{1}^{ti-1}}{(\alpha
_{1}^{t}+\alpha_{1}+1)^{i}} &
 \ldots &
\frac{\alpha _{n_{3}-1}^{ti-1}}{(\alpha _{n_{3}-1}^{t}+\alpha
_{n_{3}-1}+1)^{i}}& 0
\end{array}
\right],
 \end{equation}
where $\alpha_{n_{3}}=0$. Therefore,
$$
\frac{1}{(\alpha _{n_{3}}^{t}+\alpha _{n_{3}}+1)^{i}}=1 .
$$
It is clear that component-wise raising to the power $t$ of
the $(i+1)$-th row of  the parity check matrix gives us the next, i.e., the
$(ti+1)$-th check row:
$$
\begin{array}{r}
\left(\left[
\begin{array}{cccc}
\frac{\alpha _{1}^{i}}{(\alpha _{1}^{t}+\alpha_{1}+1)^{i}} &
 \ldots &
\frac{\alpha _{n_{3}-1}^{i}}{(\alpha _{n_{3}-1}^{t}+\alpha
_{n_{3}-1}+1)^{i}}& 0
\end{array}
\right]\right)^{t}=\\
\left[
\begin{array}{cccc}
\frac{\alpha _{1}^{ti}}{(\alpha _{1}^{t}+\alpha_{1}+1)^{i}} &
 \ldots &
\frac{\alpha _{n_{3}-1}^{ti}}{(\alpha _{n_{3}-1}^{t}+\alpha
_{n_{3}-1}+1)^{i}}& 0
\end{array}
\right] .
\end{array}
$$

The parity check matrix with the new added row can be rewritten as follows:
\begin{equation*}
\begin{array}{l}
H_{3}^{(i)}=\left[
\begin{array}{cccc}
\frac{1}{(\alpha _{1}^{t}+\alpha_{1}+1)^{i}} & \ldots &
\frac{1}{(\alpha _{n_{3}-1}^{t}+\alpha _{n_{3}-1}+1)^{i}}& 1
\\
\frac{\alpha _{1}}{(\alpha _{1}^{t}+\alpha_{1}+1)^{i}} &
 \ldots &
\frac{\alpha _{n_{3}-1}}{(\alpha _{n_{3}-1}^{t}+\alpha
_{n_{3}-1}+1)^{i}}& 0
\\
\ldots &  \ldots & \ldots &0\\
\frac{\alpha _{1}^{i}}{(\alpha _{1}^{t}+\alpha_{1}+1)^{i}} &
 \ldots &
\frac{\alpha _{n_{3}-1}^{i}}{(\alpha _{n_{3}-1}^{t}+\alpha
_{n_{3}-1}+1)^{i}}& 0\\
\ldots &  \ldots & \ldots &0\\
 \frac{\alpha _{1}^{ti-1}}{(\alpha
_{1}^{t}+\alpha_{1}+1)^{i}} &
 \ldots &
\frac{\alpha _{n_{3}-1}^{ti-1}}{(\alpha _{n_{3}-1}^{t}+\alpha
_{n_{3}-1}+1)^{i}}& 0\\
\frac{\alpha _{1}^{ti}}{(\alpha _{1}^{t}+\alpha_{1}+1)^{i}} &
 \ldots &
\frac{\alpha _{n_{3}-1}^{ti}}{(\alpha _{n_{3}-1}^{t}+\alpha
_{n_{3}-1}+1)^{i}}& 0
\end{array}
\right] = \\
\left[
\begin{array}{cccc}
1 & \ldots & 1& 1
\\
\frac{\alpha _{1}}{(\alpha _{1}^{t}+\alpha_{1}+1)^{i}} &
 \ldots &
\frac{\alpha _{n_{3}-1}}{(\alpha _{n_{3}-1}^{t}+\alpha
_{n_{3}-1}+1)^{i}}& 0
\\
\ldots &  \ldots & \ldots &0\\
\frac{\alpha _{1}^{i}}{(\alpha _{1}^{t}+\alpha_{1}+1)^{i}} &
 \ldots &
\frac{\alpha _{n_{3}-1}^{i}}{(\alpha _{n_{3}-1}^{t}+\alpha
_{n_{3}-1}+1)^{i}}& 0\\
\ldots &  \ldots & \ldots &0\\
 \frac{\alpha _{1}^{ti-1}}{(\alpha
_{1}^{t}+\alpha_{1}+1)^{i}} &
 \ldots &
\frac{\alpha _{n_{3}-1}^{ti-1}}{(\alpha _{n_{3}-1}^{t}+\alpha
_{n_{3}-1}+1)^{i}}& 0\\
\frac{\alpha _{1}^{ti}}{(\alpha _{1}^{t}+\alpha_{1}+1)^{i}} &
 \ldots &
\frac{\alpha _{n_{3}-1}^{ti}}{(\alpha _{n_{3}-1}^{t}+\alpha
_{n_{3}-1}+1)^{i}}& 0
\end{array}
\right].
\end{array}
 \end{equation*}
 Now we can obtain the following relation between the $H_{3}^{(i)} $ and $H_{3}^{*(i)}$ matrices:
 $$
H_{3}^{(i)} =\left[
\begin{array}{cc}
1  \ldots  1& 1\\
H_{3}^{*(i)}&0
\end{array}
\right].
 $$
The auxiliary $C_{3}^{*}$-code is obtained by shortening of the
$\Gamma_{3}^{(i)}$-code in a redundancy symbol corresponding to
the numerator $\{0\} \in L_{3}$. Thus, the parity check matrix
$H^{*(i)}_{3}$ of the $C_{3}^{*}$-code results from deleting
the first unit row and the last column of the parity check
matrix $H_{3}^{(i)}$ of the $\Gamma_{3}^{(i)}$-code. It follows
that $n_{3}^{*(i)}=n_{3}-1, \; k_{3}^{*(i)}=k_{3}^{(i)}, \;
d_{3}^{*(i)} \leq d_{3}^{(i)}$.
\subsection{Relation between the auxiliary $ C_{3}^{*(i)}$-code and
cumulative-separable   $\Gamma_{4}^{*(i)}$-code }
 We define that
 $$
\begin{array}{l}
\Gamma_{4}^{*(i)}=\Gamma(L_{4}^{*},G_{4}^{(i)}) \\
\text{ with } L_{4}^{*} = \left\{GF(t^{2}) \backslash
\left\{\{\alpha:
G_{4}(\alpha)=0\}\bigcup {0}\right\}\right\} ,\\
 n_{4}^{*(i)}=n_{4}^{*}
\text{ and  } G_{4}^{(i)}(x)=(x^{t}+x^{t-1}+1)^{i} .
\end{array}
$$
The $ C_{3}^{*(i)}$-code is equivalent to the
$\Gamma_{4}^{*(i)}$-code with up to the permutation
$\alpha_{i}\rightarrow \frac{1}{\alpha_{i}} , i=1, \ldots,
n_{3}-1$. Indeed, let us write the $H_{3}^{*(i)}$ and perform
the following permutation:
$$
\begin{array}{r}
H_{3}^{*(i)}= \left[
\begin{array}{ccc}

\frac{\alpha _{1}}{(\alpha _{1}^{t}+\alpha_{1}+1)^{i}} &
 \ldots &
\frac{\alpha _{n_{3}-1}}{(\alpha _{n_{3}-1}^{t}+\alpha
_{n_{3}-1}+1)^{i}}
\\
\ldots &  \ldots & \ldots \\
\frac{\alpha _{1}^{i}}{(\alpha _{1}^{t}+\alpha_{1}+1)^{i}} &
 \ldots &
\frac{\alpha _{n_{3}-1}^{i}}{(\alpha _{n_{3}-1}^{t}+\alpha
_{n_{3}-1}+1)^{i}}\\
\ldots &  \ldots & \ldots \\
 \frac{\alpha _{1}^{ti-1}}{(\alpha
_{1}^{t}+\alpha_{1}+1)^{i}} &
 \ldots &
\frac{\alpha _{n_{3}-1}^{ti-1}}{(\alpha _{n_{3}-1}^{t}+\alpha
_{n_{3}-1}+1)^{i}}\\
\frac{\alpha _{1}^{ti}}{(\alpha _{1}^{t}+\alpha_{1}+1)^{i}} &
 \ldots &
\frac{\alpha _{n_{3}-1}^{ti}}{(\alpha _{n_{3}-1}^{t}+\alpha
_{n_{3}-1}+1)^{i}}
\end{array}
\right]
\begin{array}{c}
= \\
\alpha_{i}\rightarrow \frac{1}{\alpha_{i}}
\end{array}
\\
\left[
\begin{array}{ccc}
\frac{\alpha _{1}^{ti}}{(\alpha
_{1}^{t}+\alpha_{1}^{t-1}+1)^{i}} &
 \ldots &
\frac{\alpha _{n_{3}-1}^{ti}}{(\alpha _{n_{3}-1}^{t}+\alpha
_{n_{3}-1}^{t-1}+1)^{i}}
\\
\ldots &  \ldots & \ldots \\
\frac{\alpha _{1}^{ti-i}}{(\alpha
_{1}^{t}+\alpha_{1}^{t-1}+1)^{i}} &
 \ldots &
\frac{\alpha _{n_{3}-1}^{ti-i}}{(\alpha _{n_{3}-1}^{t}+\alpha
_{n_{3}-1}^{t-1}+1)^{i}}\\
\ldots &  \ldots & \ldots \\
 \frac{\alpha _{1}}{(\alpha
_{1}^{t}+\alpha_{1}^{t-1}+1)^{i}} &
 \ldots &
\frac{\alpha _{n_{3}-1}}{(\alpha _{n_{3}-1}^{t}+\alpha
_{n_{3}-1}^{t-1}+1)^{i}}\\
\frac{1}{(\alpha _{1}^{t}+\alpha_{1}^{t-1}+1)^{i}} &
 \ldots &
\frac{1}{(\alpha _{n_{3}-1}^{t}+\alpha _{n_{3}-1}^{t-1}+1)^{i}}
\end{array}
\right] .
\end{array}
$$
 It is clear that the first row of this matrix is the $t$-th power of its last row:
$$
\begin{array}{r}
\left(\left[
\begin{array}{ccc}
\frac{1}{(\alpha _{1}^{t}+\alpha_{1}^{t-1}+1)^{i}} &
 \ldots &
\frac{1}{(\alpha _{n_{3}-1}^{t}+\alpha _{n_{3}-1}^{t-1}+1)^{i}}
\end{array}\right]\right)^{t}
=\\
\left[
\begin{array}{ccc}
\frac{\alpha _{1}^{ti}}{(\alpha
_{1}^{t}+\alpha_{1}^{t-1}+1)^{i}} &
 \ldots &
\frac{\alpha _{n_{3}-1}^{ti}}{(\alpha _{n_{3}-1}^{t}+\alpha
_{n_{3}-1}^{t-1}+1)^{i}}
\end{array}
\right] .
\end{array}
$$
It means that the above permutation in the $ C_{3}^{*(i)}$-code transforms it into the
$\Gamma_{4}^{*(i)}$-code with the parity check matrix $H_{4}^{*(i)}$:
$$
H_{4}^{*(i)}=\left[
\begin{array}{ccc}
\frac{1}{(\alpha _{1}^{t}+\alpha_{1}^{t-1}+1)^{i}} &
 \ldots &
\frac{1}{(\alpha _{n_{3}-1}^{t}+\alpha
_{n_{3}-1}^{t-1}+1)^{i}}\\
\ldots &  \ldots & \ldots \\
 \frac{\alpha _{1}^{ti-i}}{(\alpha
_{1}^{t}+\alpha_{1}^{t-1}+1)^{i}} &
 \ldots &
\frac{\alpha _{n_{3}-1}^{ti-i}}{(\alpha _{n_{3}-1}^{t}+\alpha
_{n_{3}-1}^{t-1}+1)^{i}}\\
\ldots &  \ldots & \ldots \\
 \frac{\alpha _{1}^{ti-1}}{(\alpha
_{1}^{t}+\alpha_{1}^{t-1}+1)^{i}} &
 \ldots &
\frac{\alpha _{n_{3}-1}^{ti-1}}{(\alpha _{n_{3}-1}^{t}+\alpha
_{n_{3}-1}^{t-1}+1)^{i}}
\end{array}
\right].
$$
Thus, $n_{4}^{*(i)}=n_{4}^{*}=n_{3}^{*},\; k_{4}^{*(i)}=
k_{3}^{*(i)}, \;  d_{4}^{*(i)}=d_{3}^{*(i)} $.

\subsection{Relation between the cumulative-separable codes
  $\Gamma_{4}^{*(i)}$ and  $\Gamma_{5}^{(i)}$}
We define
$$
\begin{array}{r}
\Gamma_{5}^{(i)}=\Gamma(L_{5},G_{5}^{(i)}) \text{ with } L_{5}
= L_{4}^{*},\\
 n_{5}^{(i)}=n_{5} =n_{4}^{*}\text{ and }
G_{5}^{(i)}(x)=(x^{t+1}+x^{t}+x)^{i}.
\end{array}
$$
Similarly to the case with the $\Gamma_{2}^{(i)}$-code and
$\Gamma_{1}^{*(i)}$-code, two cases characterized by
the cumulativity order $i$ should be considered:

$\boldsymbol{i < q-1 \; :}$
In this case the $\Gamma
_{5}^{(i)}$-code is  a subcode of the $\Gamma
_{4}^{*(i)}$-code.
   \begin{lemma}~\label{lemma_i<q-1_G4*_G5}
   The parity check matrices of the codes  $\Gamma_{5}^{(i)}$ and  $\Gamma_{4}^{*(i)}$
are related as:
$$
H_{5}^{(i)}= \left[
\begin{array}{ccc}
\frac{1}{G_{5}^{(i)}(\alpha _{i_{1}})} & \ldots&
\frac{1}{G_{5}^{(i)}(\alpha
_{i_{n_{5}}})} \\
& H_{4}^{*(i)} &
\end{array}
\right],
$$
where $H_{4}^{*(i)}$ is the parity check matrix of the
$\Gamma_{4}^{*(i)}$-code.

\begin{IEEEproof}

Let us use representation (\ref{parity-check_matrix_v1})
of the parity check matrix of the
$\Gamma_{5}^{(i)}$-code:
\begin{equation*}
H_{5}^{(i)}=\left[
\begin{array}{ccc}
\frac{1}{\alpha _{1}^{i}(\alpha _{1}^{t}+\alpha
_{1}^{t-1}+1)^{i}} & \ldots & \frac{1}{\alpha
_{n_{5}}^{i}(\alpha _{n_{5}}^{t}+\alpha _{n_{5}}^{t-1}+1)^{i}}
\\
\frac{1}{\alpha _{1}^{i-1}(\alpha _{1}^{t}+\alpha
_{1}^{t-1}+1)^{i}} &
 \ldots &
\frac{1}{\alpha _{n_{5}}^{i-1}(\alpha _{n_{5}}^{t}+\alpha
_{n_{5}}^{t-1}+1)^{i}}
\\
\ldots &  \ldots & \ldots \\
\frac{1}{\alpha _{1}(\alpha _{1}^{t}+\alpha _{1}^{t-1}+1)^{i}}
& \ldots & \frac{1}{\alpha _{n_{5}}(\alpha _{n_{5}}^{t}+\alpha
_{n_{5}}^{t-1}+1)^{i}}
\\
\frac{1}{(\alpha _{1}^{t}+\alpha _{1}^{t-1}+1)^{i}} &  \ldots &
\frac{1}{(\alpha _{n_{5}}^{t}+\alpha _{n_{5}}^{t-1}+1)^{i}}
\\
\ldots & \ldots  & \ldots \\
\frac{\alpha _{1}^{(i-1)t-1}}{(\alpha _{1}^{t}+\alpha
_{1}^{t-1}+1)^{i}} & \ldots& \frac{\alpha
_{n_{5}}^{(i-1)t-1}}{(\alpha _{n_{5}}^{t}+\alpha
_{n_{5}}^{t-1}+1)^{i}}
\end{array}
\right]
\end{equation*}
\begin{equation}
=  \left[
\begin{array}{ccc}
\frac{1}{\alpha _{1}^{i}(\alpha _{1}^{t}+\alpha
_{1}^{t-1}+1)^{i}} &
 \ldots & \frac{1}{\alpha _{n_{5}}^{i}(\alpha _{n_{5}}^{t}+\alpha _{n_{5}}^{t-1}+1)^{i}}
\\
\frac{1}{\alpha _{1}^{i-1}(\alpha _{1}^{t}+\alpha
_{1}^{t-1}+1)^{i}} & \ldots & \frac{1}{\alpha
_{n_{5}}^{i-1}(\alpha _{n_{5}}^{t}+\alpha
_{n_{5}}^{t-1}+1)^{i}}
\\
\ldots & \ldots  & \ldots \\
\frac{1}{\alpha _{1}(\alpha _{1}^{t}+\alpha _{1}^{t-1}+1)^{i}}
& \ldots & \frac{1}{\alpha _{n_{5}}(\alpha _{n_{5}}^{t}+\alpha
_{n_{5}}^{t-1}+1)^{i}}
\\
 & H_{4}^{*(i)} &
 \end{array} \right].
\label{parity-check_matrix_comul_G5}
\end{equation}
It is easy to see that in case of component-wise raising of any
 row
$$
\begin{array}{c}
 \left[
\begin{array}{ccc}\frac{1}{\alpha _{1}^{j}(\alpha
_{1}^{t}+\alpha _{1}^{t-1}+1)^{i}} &  \ldots & \frac{1}{\alpha
_{n_{5}}^{j}(\alpha _{n_{5}}^{t}+\alpha _{n_{5}}^{t-1}+1)^{i}}
\end{array} \right] ,\\
 0<j<i<q-1
 \end{array}
 $$
of the matrix $H_{5}^{(i)}$ to the power $t$ we obtain the
corresponding row of the submatrix $H_{4}^{*(i)}$.
$$
\begin{array}{r}
\left[
\begin{array}{ccc}\frac{1}{\alpha _{1}^{j}(\alpha
_{1}^{t}+\alpha _{1}^{t-1}+1)^{i}} &  \ldots & \frac{1}{\alpha
_{n_{5}}^{j}(\alpha _{n_{5}}^{t}+\alpha _{n_{5}}^{t-1}+1)^{i}}
\end{array} \right]^{t}=\\
\left[\begin{array}{ccc}\frac{1}{\alpha _{1}^{jt}(\alpha
_{1}+\alpha _{1}^{1-t}+1)^{i}} &  \ldots & \frac{1}{\alpha
_{n_{5}}^{jt}(\alpha _{n_{5}}+\alpha _{n_{5}}^{1-t}+1)^{i}}
\end{array} \right]=\\
\left[
\begin{array}{ccc}\frac{\alpha _{1}^{t(i-j)-i}}{(\alpha
_{1}^{t}+\alpha _{1}^{t-1}+1)^{i}} &  \ldots & \frac{\alpha
_{n_{5}}^{t(i-j)-i}}{(\alpha _{n_{5}}^{t}+\alpha
_{n_{5}}^{t-1}+1)^{i}} \end{array} \right].
\end{array}
$$
Thus, the parity check matrix of the $\Gamma_{5}^{(i)}$-code can be rewritten as
\begin{equation}
H_{5}^{(i)}=\left[
\begin{array}{cl}\frac{1}{\alpha _{1}^{i}(\alpha _{1}^{t}+\alpha _{1}^{t-1}+1)^{i}} &
 \ldots  \frac{1}{\alpha _{n_{5}}^{i}(\alpha _{n_{5}}^{t}+\alpha _{n_{5}}^{t-1}+1)^{i}}
\\
 & H_{4}^{*(i)}
 \end{array} \right], \label{parity-check_matrix_comul_G5_G4*}
\end{equation}

where the first row consists of  elements of the
$GF(t)$ only:
$$
\begin{array}{r}
\left[
\begin{array}{ccc}\frac{1}{\alpha _{1}^{i}(\alpha _{1}^{t}+\alpha _{1}^{t-1}+1)^{i}} &
 \ldots & \frac{1}{\alpha _{n_{5}}^{i}(\alpha _{n_{5}}^{t}+\alpha _{n_{5}}^{t-1}+1)^{i}}
 \end{array}
 \right]=\\
 \left[
\begin{array}{ccc}\frac{1}{(\alpha _{1}^{t+1}+\alpha _{1}^{t}+\alpha _{1})^{i}} &
 \ldots & \frac{1}{(\alpha _{n_{5}}^{t+1}+\alpha _{n_{5}}^{t}+\alpha _{n_{5}})^{i}}
 \end{array}
 \right] = \\
 \left[
\begin{array}{ccc}
\frac{1}{G_{5}^{(i)}(\alpha _{i_{1}})} & \ldots&
\frac{1}{G_{5}^{(i)}(\alpha _{i_{n_{5}}})}
\end{array}
\right].
\end{array}
$$
Hence, the redundancy of the $\Gamma_{5}^{(i)}$-code is
determined by the inequality $r_{5}^{(i)}\leq r_{4}^{*(i)} + l$
.
\end{IEEEproof}
\end{lemma}

$\boldsymbol{i =q-1 ,\;  i=q \; :}$
 In accordance with the previously  obtained Corollary \ref{equivalents_of_q-1_and_q_codes}
  and Corollary \ref{cumulative_code_as_subcode} we have  that the
    $\Gamma
    _{5}^{(q-1)}$-code is equivalent to the $\Gamma
    _{5}^{(q)}$-code and the $\Gamma _{4}^{*(q-1)}$-code is equivalent to the
$\Gamma _{4}^{*(q)}$-code.
    Let us prove now the equivalence of the codes $\Gamma_{5}^{(q-1)}$
    and $\Gamma _{4}^{*(q-1)}$.

\begin{lemma}
   The parity check matrices  of the  $\Gamma_{5}^{(q-1)}$-code
   and  $\Gamma_{4}^{*(q-1)}$-code are equal:
$$
H_{5}^{(q-1)}=  H_{4}^{*(q-1)}
$$
\begin{IEEEproof}

Using Lemma \ref{lemma_i<q-1_G4*_G5}, we write a relation
between parity check matrices of the codes $\Gamma_{5}^{(q-1)}$
and $\Gamma_{4}^{*(q-1)}$:
\begin{equation*}
H_{5}^{(q-1)}= \left[
\begin{array}{cl}
\frac{1}{\alpha _{1}^{q-1}(\alpha _{1}^{t}+\alpha
_{1}^{t-1}+1)^{q-1}} &
 \ldots \frac{1}{\alpha_{n_{5}}^{q-1}(\alpha_{n_{5}}^{t}+\alpha_{n_{5}}^{t-1}+1)^{q-1}}
\\
 & H_{4}^{*(q-1)}
 \end{array} \right].
\end{equation*}
As the codes  $\Gamma _{4}^{*(q-1)}$ and  $\Gamma _{4}^{*(q)}$
are equivalent then their parity check matrices are equal:
$H_{4}^{*(q-1)}= H_{4}^{*(q)} $.
We show that the row $$\left[
\begin{array}{ccc}
\frac{1}{\alpha _{1}^{q-1}(\alpha _{1}^{t}+\alpha
_{1}^{t-1}+1)^{q-1}} &
 \ldots & \frac{1}{\alpha _{n_{5}}^{q-1}(\alpha _{n_{5}}^{t}+\alpha _{n_{5}}^{t-1}+1)^{q-1}}\end{array} \right]$$
 of the matrix  $H_{5}^{(q-1)}$
 can be obtained as a linear combination of the matrix   $H_{4}^{*(q)}$ row:
 $$\left[\begin{array}{ccc}
\frac{\alpha _{1}^{t-q}}{(\alpha _{1}^{t}+\alpha
_{1}^{t-1}+1)^{q}} &
 \ldots & \frac{\alpha _{n_{5}}^{t-q}}{(\alpha _{n_{5}}^{t}+\alpha _{n_{5}}^{t-1}+1)^{q}}\end{array}
 \right],$$
 its $t$-th power:
$$
\begin{array}{c}
\left(\left[
\begin{array}{ccc}
\frac{\alpha _{1}^{t-q}}{(\alpha _{1}^{t}+\alpha
_{1}^{t-1}+1)^{q}} &
 \ldots & \frac{\alpha _{n_{5}}^{t-q}}{(\alpha _{n_{5}}^{t}+
 \alpha _{n_{5}}^{t-1}+1)^{q}}\end{array} \right]\right)^{t}=\\
 \left[
\begin{array}{ccc}
\frac{1}{\alpha _{1}^{q-1}(\alpha _{1}^{t}+\alpha
_{1}^{t-1}+1)^{q}} &
 \ldots & \frac{1}{\alpha _{n_{5}}^{q-1}(\alpha _{n_{5}}^{t}+\alpha _{n_{5}}^{t-1}+1)^{q}}\end{array}
 \right]
\end{array}
$$
and the matrix  $H_{4}^{*(q)}$ row:
$$
\left[
\begin{array}{ccc}
\frac{\alpha _{1}^{t-q+1}}{(\alpha _{1}^{t}+\alpha
_{1}^{t-1}+1)^{q}} &
 \ldots& \frac{\alpha _{n_{5}}^{t-q+1}}{(\alpha _{n_{5}}^{t}+
 \alpha _{n_{5}}^{t-1}+1)^{q}}\end{array} \right].
$$
 Indeed,

$$\begin{array}{r}

\left[
\begin{array}{ccc}
\frac{\alpha _{1}^{t-q+1}}{(\alpha _{1}^{t}+\alpha
_{1}^{t-1}+1)^{q}} &
 \ldots & \frac{\alpha _{n_{5}}^{t-q+1}}{(\alpha _{n_{5}}^{t}+
 \alpha _{n_{5}}^{t-1}+1)^{q}}\end{array} \right]+\\
\left[
\begin{array}{ccc}
\frac{1}{\alpha _{1}^{q-1}(\alpha _{1}^{t}+\alpha
_{1}^{t-1}+1)^{q}} & \ldots & \frac{1}{\alpha
_{n_{5}}^{q-1}(\alpha _{n_{5}}^{t}+\alpha
_{n_{5}}^{t-1}+1)^{q}}\end{array}
 \right] +\\
\left[\begin{array}{ccc} \frac{\alpha _{1}^{t-q}}{(\alpha
_{1}^{t}+\alpha _{1}^{t-1}+1)^{q}} &
 \ldots & \frac{\alpha _{n_{5}}^{t-q}}{(\alpha _{n_{5}}^{t}+\alpha _{n_{5}}^{t-1}+1)^{q}}\end{array}
 \right] =\\
 \left[
\begin{array}{ccc}
\frac{\alpha _{1}^{t}+\alpha _{1}^{t-1}+1}{\alpha
_{1}^{q-1}(\alpha _{1}^{t}+\alpha _{1}^{t-1}+1)^{q}} &
 \ldots & \frac{\alpha _{n_{5}}^{t}+\alpha _{n_{5}}^{t-1}+1}{\alpha _{n_{5}}^{q-1}(\alpha _{n_{5}}^{t}+\alpha _{n_{5}}^{t-1}+1)^{q}}\end{array}
 \right] =\\
 \left[
\begin{array}{ccc}
\frac{1}{\alpha _{1}^{q-1}(\alpha _{1}^{t}+\alpha
_{1}^{t-1}+1)^{q-1}} &
 \ldots & \frac{1}{\alpha _{n_{5}}^{q-1}(\alpha _{n_{5}}^{t}+\alpha _{n_{5}}^{t-1}+1)^{q-1}}
 \end{array}
 \right]
 \end{array}
$$

\end{IEEEproof}
\end{lemma}
Hence, the redundancy of the $\Gamma_{5}^{(q-1)}$-code is equal
to that of the $\Gamma_{4}^{*(q-1)}$-code:
$$r_{5}^{(q-1)}= r_{4}^{*(q-1)} .$$
\subsection{Relation between the cumulative-separable codes $\Gamma_{5}^{(i)}$
and  $\Gamma_{6}^{(i)}$}

  We define the $\Gamma_{6}^{(i)}$-code as:
  $$
  \begin{array}{r}
  \Gamma_{6}^{(i)}=\Gamma(L_{6},G_{6}^{(i)}) \text{ with } L_{6} =
\{GF(t^{2}) \backslash \{\alpha : G_{6}(\alpha)=0\}\} ,\\
n_{6}^{(i)}=n_{6} \text{ and } G_{6}^{(i)}(x)=(x^{t+1}+1)^{i}
 \end{array}$$
It is clear that the codes
$\Gamma_{6}^{(i)}$ and $\Gamma _{5}^{(i)}$ are  equivalent for all
$i\geq 1$ up to the permutation
\begin{center}
$ \beta \rightarrow \gamma\beta-1 \text{ , where }
    \gamma \in GF(t^{2}) \text{ and } \gamma^{t+1}+1=0 \text{ , } \beta \in L_{5} .
$
\end{center}
and therefore, $n_{6}=n_{5}$ , $k_{6}^{(i)}=k_{5}^{(i)}$ and
$d_{6}^{(i)}=d_{5}^{(i)}$.

\begin{figure*}[!H]
\center
\includegraphics[width=0.6\textwidth, angle=-90]{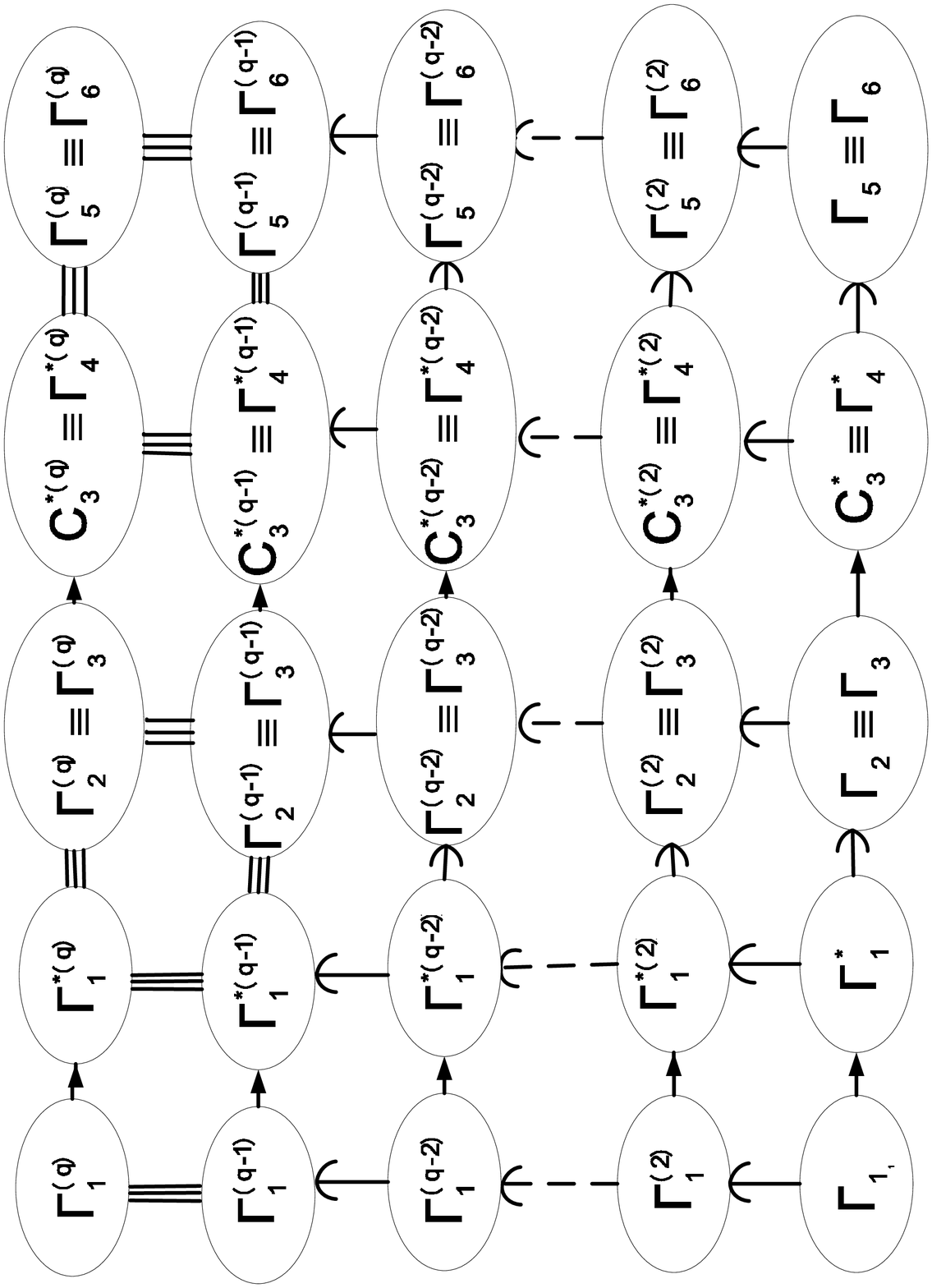}
\caption{Embedded and equivalent cumulative-separable  codes }
\label{fig1_cumulativeseparable}
\end{figure*}

 Thus, all the relations considered above and together with Corollary
\ref{subcodes_in_comul_separable_codes} allow us  to present
the following Table
\ref{tab:Table_of_Comulative_Separable_codes} and pattern in
Fig. \ref{fig1_cumulativeseparable}.

\begin{table*}[!H]
\caption{Cumulative-separable codes}

\begin{tabular}{|c|c|c|c|c|c|}
\hline  $i$& $  \begin{array}{c}
   \Gamma_{1}^{(i)}\\
   n_{1}=n^{*}_{1}+1

 \end{array}
 $
&
 $ \begin{array}{c}
   \Gamma_{1}^{*(i)}\\
  n^{*}_{1}=n_{2}
 \end{array}
 $
&
 $ \begin{array}{c}
   \Gamma^{(i)}_{2} \equiv \Gamma^{(i)}_{3}\\
 n_{2}=n_{3}=n^{*}_{3}+1
 \end{array}$
&
 $ \begin{array}{c}
   C^{*(i)}_{3} \equiv \Gamma^{*(i)}_{4}  \\
  n^{*}_{3}=n^{*}_{4}=n_{5},\\
\end{array}$
&
 $  \begin{array}{c}
 \Gamma^{(i)}_{5} \equiv \Gamma^{(i)}_{6} \\
  n_{5}=n_{6}=t^{2}-t-1,\\
  \end{array}$
 \\ \hline
$ \leq q-2$ & $\begin{array}{c}
k^{(i)}_{1}=k^{*(i)}_{1}+1 ,\\
d^{(i)}_{1}=i(t-1)+1
 \end{array}$&
 $\begin{array}{c}
 k^{*(i)}_{1} \leq k^{(i)}_{2}+l,\\
d^{(i)}_{1} \leq d^{*(i)}_{1} \leq d^{(i)}_{2}
 \end{array}$
 &
$\begin{array}{c}
 k^{(i)}_{2}=k^{(i)}_{3}=k^{*(q)}_{3},\\
 d^{(i)}_{2}=d^{(i)}_{3} \geq d^{*(i)}_{3}
 \end{array}$
  &
$\begin{array}{c}
 k^{*(i)}_{3}= k^{*(i)}_{4} \leq k^{(i)}_{5}+l,\\
  d^{*(i)}_{3}=d^{*(i)}_{4} \leq d^{(i)}_{5}
 \end{array}$
   &
$\begin{array}{c}
 k^{(i)}_{5} = k^{(i)}_{6},\\
 k^{(i)}_{6} \geq t^{2}-t-1-\\
2lj\left(t+1-\frac{j+3}{2}\right),\\
d^{(i)}_{5}=d^{(i)}_{6}=i(t+1)+1
\end{array}$
 \\ \hline
$ q , \; q-1$ & $\begin{array}{c}
k^{(q)}_{1}=k^{*(q)}_{1}+1 ,\\
d^{(q)}_{1} \geq q(t-1)+1
 \end{array}$
 &
$ \begin{array}{c}
 k^{*(q)}_{1}=k^{(q)}_{2},\\
d^{(q)}_{1} \leq d^{*(q)}_{1}=d^{(q)}_{2}
 \end{array}$
 &
$\begin{array}{c}
k^{(q)}_{2}=k^{(q)}_{3}= k^{*(q)}_{3},\\
 d^{(q)}_{2}=d^{(q)}_{3}\geq d^{*(q)}_{3}
 \end{array}$
  &
$\begin{array}{c}
 k^{*(q)}_{3}= k^{*(q)}_{4}=k^{(q)}_{5},\\
 d^{*(q)}_{3}=d^{*(q)}_{4}= d^{(q)}_{5}
 \end{array}$
   &
$\begin{array}{c}
 k^{(q)}_{5} = k^{(q)}_{6},\\
 k^{(q)}_{6}\geq
t^{2}-t-1-\\
2l(q-1)\left(t-\frac{3}{2}-\frac{q-2}{2}\right),\\
 d^{(q)}_{5}=d^{(q)}_{6} \geq q(t+1)+1
\end{array}$
 \\ \hline
\end{tabular}
\label{tab:Table_of_Comulative_Separable_codes}
\end{table*}

\section{Minimum distance of cumulative-separable codes }
The minimum distances of primary separable codes $\Gamma_{1}$ -
$\Gamma_{6}$ have been determined in \cite{Bezz7,Bezz10}. Now
let us take the $\Gamma_{6}$-code as an example to determine
the minimum distances of the corresponding separable-cumulative
codes $\Gamma_{6}^{(i)}$  in case of the cumulativity order
$i>1$.

Let us consider $q-1$  groups of elements that form a set
$L_{6}$:

\textbf{1. }
$L_{6(1)}^{(j)}=\left\{\alpha^{(t-1)l+\frac{t-1}{q-1}j}
\right\}_{l=0,\ldots,t} ; j=0,\ldots,q-2$
   with the exception of $j=j^{*}$ , such that
   $$G_{6}(\alpha^{(t-1)l+\frac{t-1}{q-1}j^{*}})=\alpha^{\frac{(t-1)(t+1)}{q-1}j^{*}}+1=0 .$$
   In other words,  we exclude $j=j^{*}$  such that
   $$ \alpha^{\frac{(t-1)(t+1)}{q-1}j^{*}}=q-1 ,
\alpha \text{ is a primitive element of }  GF(t^{2}) .$$
 It is
easy to show that
$$
L_{6(1)}^{(i)} \cap L_{6(1)}^{(j)}=\emptyset \text{  for any }
i \neq j  ; \; i,j=0,\ldots,q-2 .
$$

 Therefore the number of different groups $L_{6(1)}^{(j)}$ in $L_{6}$ is $q-2$.
Further, we shall denote  vector \textbf{a} values in the set
$L_{6(1)}^{(j)}$ positions by
$$
\begin{array}{rl}
a_{j,l,1} \; , & j=0,1,\ldots,j^{*}-1,j^{*}+1,\ldots,q-2;\\
 &l=0,\ldots,t \; .
 \end{array}
 $$
\textbf{2. }$L_{6(0)}=\left\{0\right\}$. The vector \textbf{a}
value in this position is denoted by $a_{0}$.

It should be noted that the value of the polynomial $G_{6}(x)$
for all  elements from one numerator group is  the same and it
is equal to:

\textbf{1. }For elements from the numerator group
$L_{6(1)}^{(j)}$ :
  $$
  \begin{array}{r}
  G_{6}^{(i)}(\alpha^{(t-1)l+\frac{t-1}{q-1}j})=\left((\alpha^{(t-1)l+\frac{t-1}{q-1}j})^{t+1}+1\right)^{i}
  =\\
  \left(\alpha^{\frac{(t-1)(t+1)}{q-1}j}+1\right)^{i}.
   \end{array}
 $$
Obviously, all values
$$
\begin{array}{l}
G_{6}^{(i)}(\alpha^{(t-1)l+\frac{t-1}{q-1}j}) \in GF(q) ,\; \\
 j=0,1,\ldots,j^{*}-1,j^{*}+1,\ldots,q-2 .
\end{array}
$$
\textbf{2. }  For elements from the numerator group $L_{6(0)}$:
  $G_{6}^{(i)}(0)=1.$

 Let us consider now the submatrix $H_{6(1,j)}^{(i)}$
of the parity check matrix $H_{6}^{(i)}$. It corresponds to the
numerator group  $L_{6(1)}^{(j)}$ that we determined above.
Here we use representation \eqref{parity-check_matrix_v2}:
\begin{equation}
H_{6(1,j)}^{(i)}= \left[
\begin{array}{ccc}
\frac{1}{\left(\alpha^{\frac{(t-1)(t+1)}{q-1}j}+1\right)^{i}}&
 \ldots &
\frac{1}{\left(\alpha^{\frac{(t-1)(t+1)}{q-1}j}+1\right)^{i}}
\\
\frac{\alpha^{\frac{t-1}{q-1}j}}{\left(\alpha^{\frac{(t-1)(t+1)}{q-1}j}+1\right)^{i}}
&
\ldots & \frac{\alpha^{\frac{t-1}{q-1}j+(t-1)t}}{\left(\alpha^{\frac{(t-1)(t+1)}{q-1}j}+1\right)^{i}} \\
\ldots & \ldots &  \ldots \\
\frac{\left(\alpha^{\frac{t-1}{q-1}j}\right)^{t}}{\left(\alpha^{\frac{(t-1)(t+1)}{q-1}j}+1\right)^{i}}
& \ldots &
\frac{\left(\alpha^{\frac{t-1}{q-1}j+(t-1)t}\right)^{t}}{\left(\alpha^{\frac{(t-1)(t+1)}{q-1}j}+1\right)^{i}}\\
\frac{1}{\left(\alpha^{\frac{(t-1)(t+1)}{q-1}j}+1\right)^{i-1}}&
\ldots &
\frac{1}{\left(\alpha^{\frac{(t-1)(t+1)}{q-1}j}+1\right)^{i-1}}
\\
\frac{\alpha^{\frac{t-1}{q-1}j}}{\left(\alpha^{\frac{(t-1)(t+1)}{q-1}j}+1\right)^{i-1}}
&
\ldots & \frac{\alpha^{\frac{t-1}{q-1}j+(t-1)t}}{\left(\alpha^{\frac{(t-1)(t+1)}{q-1}j}+1\right)^{i-1}} \\
\ldots& \ldots & \ldots \\
\frac{\left(\alpha^{\frac{t-1}{q-1}j}\right)^{t}}{\left(\alpha^{\frac{(t-1)(t+1)}{q-1}j}+1\right)^{2}}
& \ldots &
\frac{\left(\alpha^{\frac{t-1}{q-1}j+(t-1)t}\right)^{t}}{\left(\alpha^{\frac{(t-1)(t+1)}{q-1}j}+1\right)^{2}}\\
\frac{1}{\alpha^{\frac{(t-1)(t+1)}{q-1}j}+1}& \ldots &
\frac{1}{\alpha^{\frac{(t-1)(t+1)}{q-1}j}+1}
\\
\ldots & \ldots &  \ldots\\
\frac{\left(\alpha^{\frac{t-1}{q-1}j}\right)^{t}}{\alpha^{\frac{(t-1)(t+1)}{q-1}j}+1}
& \ldots &
\frac{\left(\alpha^{\frac{t-1}{q-1}j+(t-1)t}\right)^{t}}{\alpha^{\frac{(t-1)(t+1)}{q-1}j}+1}
\end{array}
\right]. \label{parity-check_matrix_H6_1}
\end{equation}
Let us take the matrix column corresponding to the position
$\{0\}$:
\begin{equation}
 H_{6(0)}^{(i)}=\left[
\begin{array}{c}
1\\
0\\
\vdots \\
0\\
1\\
0\\
\vdots\\
0\\
1\\
\vdots \\
0
\end{array}
\right]. \label{parity-check_matrix_H6_0}
\end{equation}

Now we write the parity check matrix of the
$\Gamma_{6}^{(i)}$-code by using the submatrices
$H_{6(1,j)}^{(i)}$ and $H_{6(0)}^{(i)}$:
\begin{center}
$ H_{6}^{(i)}= \left[
\begin{array}{ccccc}
H_{6(1,0)}^{(i)}& H_{6(1,1)}^{(i)}&\ldots&H_{6(1,q-2)}^{(i)}&
H_{6(0)}^{(i)}
\end{array}
\right] .$
\end{center}
Let us choose the vector  \textbf{a} of the following form:
$$\begin{array}{r}
 \textbf{a}=\left[
 \begin{array}{ccc}
a_{0,0,1} a_{0,1,1}\ldots a_{0,t,1}&a_{1,0,1} a_{1,1,1}\ldots
a_{1,t,1}&\ldots \end{array} \right. \\
 \left.
 \begin{array}{ccc}
 &a_{q-2,0,1} a_{q-2,1,1}\ldots a_{q-2,t,1}&a_{0}
\end{array}
\right].
\end{array}
$$
Assume that $$\begin{array}{c}
 a_{0,0,1}= a_{0,1,1}=\ldots=
a_{0,t,1}= a_{0,1} \in GF(q) ;\\
a_{1,0,1} = a_{1,1,1}=\ldots= a_{1,t,1}=a_{1,1} \in GF(q);\\
\ldots\\
a_{q-2,0,1}= a_{q-2,1,1}=\ldots =a_{q-2,t,1}= a_{q-2,t,1} \in
GF(q).
\end{array}
$$
Now we calculate the result of multiplication of the vector
\textbf{a} by the parity check matrix of the code:
\begin{equation}
 \begin{array}
{l}
 \textbf{a}\cdot H_{6}^{(i)T}=
 \left[
 a_{0,1} \ldots a_{0,1} \ldots a_{q-2,1} \ldots a_{q-2,1} a_{0}
\right]\cdot \\\left[ H_{6(1,0)}^{(i)} \ldots
H_{6(1,q-2)}^{(i)} H_{6(0)}^{(i)} \right]^{T}= \\ \left[
a_{0,1} \ldots a_{0,1}\right]\cdot\left[
H_{6(1,0)}^{(i)}\right]^{T}+ \ldots
\\+ \left[ a_{q-2,1} \ldots a_{q-2,1} \right]\cdot\left[
H_{6(1,q-2)}^{(i)}\right]^{T} +
\\
a_{0}\cdot\left[H_{6(0)}^{(i)}\right]^{T}= a_{0,1}\left[ 1
\ldots 1\right]\cdot\left[ H_{6(1,0)}^{(i)}\right]^{T}+ \ldots+
\\a_{q-2,1}\left[1 \ldots 1 \right]\cdot\left[
H_{6(1,q-2)}^{(i)}\right]^{T}+a_{0}\cdot\left[H_{6(0)}^{(i)}\right]^{T}\;
.
\end{array}\label{aH_for_minimal_distance_comulative_codes}
\end{equation}
Let us calculate the value of the row vector that is obtained
from multiplying the identity  vector by the submatrix
$H_{6(1,j)}^{(i)}$  \eqref{parity-check_matrix_H6_1}, for all
$j$ from $0$ to $q-2$ with an exception $j=j^{*}:
\alpha^{\frac{(t-1)(t+1)}{q-1}j^{*}}= q-1$:
\begin{equation*}
 \left[1 \ldots 1  1 \right]\cdot \\
 \left[
\begin{array}{ccc}
\frac{1}{\left(\alpha^{\frac{(t-1)(t+1)}{q-1}j}+1\right)^{i}}&
 \ldots &
\frac{1}{\left(\alpha^{\frac{(t-1)(t+1)}{q-1}j}+1\right)^{i}}
\\
\frac{\alpha^{\frac{t-1}{q-1}j}}{\left(\alpha^{\frac{(t-1)(t+1)}{q-1}j}+1\right)^{i}}
&
\ldots & \frac{\alpha^{\frac{t-1}{q-1}j+(t-1)t}}{\left(\alpha^{\frac{(t-1)(t+1)}{q-1}j}+1\right)^{i}} \\
\ldots & \ldots &  \ldots \\
\frac{\left(\alpha^{\frac{t-1}{q-1}j}\right)^{t}}{\left(\alpha^{\frac{(t-1)(t+1)}{q-1}j}+1\right)^{i}}
& \ldots &
\frac{\left(\alpha^{\frac{t-1}{q-1}j+(t-1)t}\right)^{t}}{\left(\alpha^{\frac{(t-1)(t+1)}{q-1}j}+1\right)^{i}}\\
\frac{1}{\left(\alpha^{\frac{(t-1)(t+1)}{q-1}j}+1\right)^{i-1}}&
\ldots &
\frac{1}{\left(\alpha^{\frac{(t-1)(t+1)}{q-1}j}+1\right)^{i-1}}
\\
\frac{\alpha^{\frac{t-1}{q-1}j}}{\left(\alpha^{\frac{(t-1)(t+1)}{q-1}j}+1\right)^{i-1}}
&
\ldots & \frac{\alpha^{\frac{t-1}{q-1}j+(t-1)t}}{\left(\alpha^{\frac{(t-1)(t+1)}{q-1}j}+1\right)^{i-1}} \\
\ldots& \ldots & \ldots \\
\frac{\left(\alpha^{\frac{t-1}{q-1}j}\right)^{t}}{\left(\alpha^{\frac{(t-1)(t+1)}{q-1}j}+1\right)^{2}}
& \ldots &
\frac{\left(\alpha^{\frac{t-1}{q-1}j+(t-1)t}\right)^{t}}{\left(\alpha^{\frac{(t-1)(t+1)}{q-1}j}+1\right)^{2}}\\
\frac{1}{\alpha^{\frac{(t-1)(t+1)}{q-1}j}+1}& \ldots &
\frac{1}{\alpha^{\frac{(t-1)(t+1)}{q-1}j}+1}
\\
\ldots & \ldots &  \ldots\\
\frac{\left(\alpha^{\frac{t-1}{q-1}j}\right)^{t}}{\alpha^{\frac{(t-1)(t+1)}{q-1}j}+1}
& \ldots &
\frac{\left(\alpha^{\frac{t-1}{q-1}j+(t-1)t}\right)^{t}}{\alpha^{\frac{(t-1)(t+1)}{q-1}j}+1}
\end{array}
\right]^{T}
\end{equation*}

\begin{equation*}
= \left[
\begin{array}{c}
\frac{1+1+\ldots
+1}{\left(\alpha^{\frac{(t-1)(t+1)}{q-1}j}+1\right)^{i}}
\\
\frac{\alpha^{\frac{t-1}{q-1}j}+\alpha^{\frac{t-1}{q-1}j+(t-1)}+
\ldots+\alpha^{\frac{t-1}{q-1}j+(t-1)t}}
{\left(\alpha^{\frac{(t-1)(t+1)}{q-1}j}+1\right)^{i}}\\
\vdots \\
\frac{\left(\alpha^{\frac{t-1}{q-1}j}\right)^{t}+
\left(\alpha^{\frac{t-1}{q-1}j+(t-1)}\right)^{t}+\ldots+
\left(\alpha^{\frac{t-1}{q-1}j+(t-1)t}\right)^{t}}
{\left(\alpha^{\frac{(t-1)(t+1)}{q-1}j}+1\right)^{i}}\\
\frac{1+1+\ldots+1}{\left(\alpha^{\frac{(t-1)(t+1)}{q-1}j}+1\right)^{i-1}}\\
\frac{\alpha^{\frac{t-1}{q-1}j}+\alpha^{\frac{t-1}{q-1}j+(t-1)}+
\ldots+\alpha^{\frac{t-1}{q-1}j+(t-1)t}}
{\left(\alpha^{\frac{(t-1)(t+1)}{q-1}j}+1\right)^{i-1}}\\
\vdots\\
\frac{\left(\alpha^{\frac{t-1}{q-1}j}\right)^{t}+
\left(\alpha^{\frac{t-1}{q-1}j+(t-1)}\right)^{t}+\ldots+
\left(\alpha^{\frac{t-1}{q-1}j+(t-1)t}\right)^{t}}
{\left(\alpha^{\frac{(t-1)(t+1)}{q-1}j}+1\right)^{2}}\\
\frac{1+1+\ldots+1}{\alpha^{\frac{(t-1)(t+1)}{q-1}j}+1}\\
\vdots \\
\frac{\left(\alpha^{\frac{t-1}{q-1}j}\right)^{t}+
\left(\alpha^{\frac{t-1}{q-1}j+(t-1)}\right)^{t}+\ldots+
\left(\alpha^{\frac{t-1}{q-1}j+(t-1)t}\right)^{t}}
{\alpha^{\frac{(t-1)(t+1)}{q-1}j}+1}
\end{array}
\right]^{T} =
\end{equation*}

\begin{equation*}
\left[
\begin{array}{c}
\frac{t+1}{\left(\alpha^{\frac{(t-1)(t+1)}{q-1}j}+1\right)^{i}}
\\
\frac{\alpha^{\frac{t-1}{q-1}j}}
{\left(\alpha^{\frac{(t-1)(t+1)}{q-1}j}+1\right)^{i}}
\left(1+\alpha^{t-1}+\ldots+\alpha^{(t-1)t}\right)
\\
\vdots\\
\frac{\left(\alpha^{\frac{t-1}{q-1}j}\right)^{t}}{\left(\alpha^{\frac{(t-1)(t+1)}{q-1}j}+1\right)^{i}}
\left(1+(\alpha^{t-1})^{t}+\ldots+(\alpha^{(t-1)t})^{t}\right)
\\
\frac{t+1}{\left(\alpha^{\frac{(t-1)(t+1)}{q-1}j}+1\right)^{i-1}}
\\
\frac{\alpha^{\frac{t-1}{q-1}j}}{\left(\alpha^{\frac{(t-1)(t+1)}{q-1}j}+1\right)^{i-1}}
\left(1+\alpha^{t-1}+\ldots+\alpha^{(t-1)t}\right)
\\
\vdots\\
\frac{\left(\alpha^{\frac{t-1}{q-1}j}\right)^{t}}{\left(\alpha^{\frac{(t-1)(t+1)}{q-1}j}+1\right)^{2}}
\left(1+(\alpha^{t-1})^{t}+\ldots+(\alpha^{(t-1)t})^{t}\right)
\\
\frac{t+1}{\alpha^{\frac{(t-1)(t+1)}{q-1}j}+1}
\\
 \vdots
\\
\frac{\left(\alpha^{\frac{t-1}{q-1}j}\right)^{t}}{\alpha^{\frac{(t-1)(t+1)}{q-1}j}+1}
\left(1+(\alpha^{t-1})^{t}+\ldots+(\alpha^{(t-1)t})^{t}\right)
\end{array}
\right]^{T}
\end{equation*}

Since $1,\alpha^{t-1},\ldots,\alpha^{(t-1)t}$ are all the roots
of the polynomial $x^{t+1}-1$ , then the sum of any nonzero
powers of these elements is always equal to zero:
$$
\sum \limits _{i=0}^{t}\alpha^{(t-1)i}=0.
$$
Besides  $t+1=1 \mod q$, and we obtain:
\begin{equation*}
 \left[1 \ldots 1  1 \right]\cdot H_{6}^{(i)T}=\left[
 \begin{array}{c}

 \frac{1}{\left(\alpha^{\frac{(t-1)(t+1)}{q-1}j}+1\right)^{i}}\\
0\\
\vdots \\
0\\
 \frac{1}{\left(\alpha^{\frac{(t-1)(t+1)}{q-1}j}+1\right)^{i-1}}\\
 0\\
 \vdots\\
 0\\
  \frac{1}{\alpha^{\frac{(t-1)(t+1)}{q-1}j}+1}\\
 \vdots\\
0
 \end{array}
 \right]^{T}
\end{equation*}

Therefore expression
\eqref{aH_for_minimal_distance_comulative_codes} can be
rewritten as:
\begin{equation}
\begin{array}{l}

\boldsymbol{a}^{*}\cdot H_{6}^{*(i)T}=
 \left[
 \begin{array}{cccc}
a_{0,1}&\ldots&a_{q-2,1}&a_{0}
\end{array}
\right]\cdot \\
\left[
\begin{array}{cccc}
\frac{1}{2^{i}}&
\ldots&\frac{1}{\left(\alpha^{\frac{(t-1)(t+1)}{q-1}t}+1\right)^{i}}& 1\\
\frac{1}{2^{i-1}}&
 \ldots& \frac{1}{\left(\alpha^{\frac{(t-1)(t+1)}{q-1}t}+1\right)^{i-1}} &1\\
\ldots&\ldots&\ldots&\ldots\\
\frac{1}{2}&
\ldots & \frac{1}{\alpha^{\frac{(t-1)(t+1)}{q-1}t}+1}& 1\\
\end{array}
\right]^{T}
\end{array}, \label{RScode_parity_check_matrix_G6_q}
\end{equation}
where the length of vector $\boldsymbol{a}^{*}$ is equal to
$q-1$.

It is easy to note that all $q-1$ different nonzero elements of
the field $GF(q)$ are in the last row of the matrix. The matrix
in \eqref{RScode_parity_check_matrix_G6_q} itself is the
Vandermonde matrix. These remarks permit us to formulate the
following theorem on the minimum distance of
cumulative-separable codes from the $\Gamma_{6}^{(i)}$ subclass
of codes.

\begin{theorem}~\label{theorem_minimal_distance_comulative_G6_q}
The minimum distance of the codes from the subclass  $\Gamma
_{6}^{(i)}$ is equal to $i(t+1)+1$ for $1<i<q-1$ .

\begin{IEEEproof}

First of all, it should be noted that  the matrix
$H_{6}^{*(i)}$ in \eqref{RScode_parity_check_matrix_G6_q} is
the parity check matrix of RS-code. It means that for $i<q-1$
there always exists a codeword of RS-code with nonzero symbol
on the position $\{0\}$ and with other $i$ nonzero symbols in
positions that correspond to  the  other columns of the parity
check matrix $H_{6}^{*(i)}$ in
\eqref{RScode_parity_check_matrix_G6_q}. Using expression
\eqref{aH_for_minimal_distance_comulative_codes} the following
value of the minimum distance of the $\Gamma_{6}^{(i)}$-code
can be obtained for $1<i<q-1$ :
\begin{equation*}
d_{6}^{(i)}=i(t+1)+1= \deg G_{6}^{(i)}(x) +1  .
\end{equation*}
\end{IEEEproof}
\end{theorem}

\section{Dimension of the cumulative-separable codes that are obtained
from the $\Gamma_{6}$-code}

According to the definition of the $\Gamma_{6}$-code the
cumulative-separable $\Gamma(L_{6},G_{6}^{(q)})$-code
corresponding to it is given by the polynomial $G_{6}^{(q)}(x)
=\left(x^{t+1}+1\right)^{q}$  and set $L_{6}=\left\{GF(t^{2})
\backslash \left\{\alpha : G_{6}(\alpha)=0\right\} \right\}$,
where $t=q^{l}$. The parity check matrix $H_{6}$ of the
$\Gamma(L_{6},G_{6}^{(q)})$-code can be presented as set of
submatrices $h_{1}, h_{2}, \ldots, h_{q-1}$ by using expression
\eqref{parity-check_matrix_v2}. Here for simplicity we
 denote the length of  the $\Gamma_{6}$ - code by $n$. \\
\textbf{1.} Consider subsets of submatrix $h_{1}$ rows that can
be obtained by  transformations of other rows of this
submatrix:
\begin{equation}
h_{1}=\left[ \begin{array}{ccccc} \frac{1}{\alpha _{1}^{t+1}+1}
& \frac{1}{\alpha _{2}^{t+1}+1}&
 \ldots & \frac{1}{\alpha _{n-1}^{t+1}+1}& 1 \\
\frac{\alpha _{1}}{\alpha _{1}^{t+1}+1} & \frac{\alpha
_{2}}{\alpha _{2}^{t+1}+1}&\ldots &
\frac{\alpha _{n-1}}{\alpha _{n-1}^{t+1}+1}& 0 \\
\ldots & \ldots & \ldots & \ldots & \ldots\\
\frac{\alpha _{1}^{t}}{\alpha _{1}^{t+1}+1} & \frac{\alpha
_{2}^{t}}{\alpha _{2}^{t+1}+1}&\ldots &
\frac{\alpha _{n-1}^{t}}{\alpha _{n-1}^{t+1}+1}& 0 \\
\end{array}
\right], \label{parity-check_matrix 1 G6_v1}
\end{equation}
where $\alpha_{i}\in L_{6} $ and $\alpha_{n}=0$.\\
\textbf{1.1.} All elements of the first row in the submatrix
$h_{1}$ are elements of the field $GF(t)$ . Indeed,
\begin{center}
$ \left(\frac{1}{\alpha _{i}^{t+1}+1}
\right)^{t}=\frac{1}{\alpha _{i}^{t+1}+1} $
\end{center}
for any $i=1,\ldots,n-1$.\\
\textbf{1.2.} The row
\begin{center}
$ \left[ \begin{array}{ccccc} \frac{\alpha _{1}^{t}}{\alpha
_{1}^{t+1}+1} & \frac{\alpha _{2}^{t}}{\alpha _{2}^{t+1}+1}&
 \ldots & \frac{\alpha _{n-1}^{t}}{\alpha _{n-1}^{t+1}+1}& 0 \\
\end{array}
\right] $
\end{center}
is obtained by element-wise raising to the power $t$ of the row
\begin{center}
$ \left[ \begin{array}{ccccc}
 \frac{\alpha _{1}}{\alpha _{1}^{t+1}+1} &
 \frac{\alpha _{2}}{\alpha _{2}^{t+1}+1}&\ldots &
 \frac{\alpha _{n-1}}{\alpha _{n-1}^{t+1}+1}& 0
\end{array}
\right]
 $
\end{center}
of the submatrix.

Therefore, the number of $q$-ary rows equal to $\delta_{1} =l+2l$
can be deleted from the submatrix $h_{1}$  as they are linearly
dependent on the other rows of this submatrix. \\
\textbf{2. }Consider now a submatrix $h_{2}$:
\begin{equation}
h_{2}=\left[ \begin{array}{ccccc}
\frac{1}{(\alpha _{1}^{t+1}+1)^{2}} &
 \frac{1}{(\alpha _{2}^{t+1}+1)^{2}}& \ldots &
 \frac{1}{(\alpha _{n-1}^{t+1}+1)^{2}}& 1 \\
\frac{\alpha _{1}}{(\alpha _{1}^{t+1}+1)^{2}} & \frac{\alpha
_{2}}{(\alpha _{2}^{t+1}+1)^{2}}&\ldots &
\frac{\alpha _{n-1}}{(\alpha _{n-1}^{t+1}+1)^{2}}& 0 \\
\ldots & \ldots & \ldots & \ldots & \ldots\\
\frac{\alpha _{1}^{t}}{(\alpha _{1}^{t+1}+1)^{2}} &
 \frac{\alpha _{2}^{t}}{(\alpha _{2}^{t+1}+1)^{2}}&\ldots &
  \frac{\alpha _{n-1}^{t}}{(\alpha _{n-1}^{t+1}+1)^{2}}& 0
\end{array}
\right], \label{parity-check_matrix 2 G6_v1}
\end{equation}
where $\alpha_{i}\in L_{6} $ and $\alpha_{n}=0$.\\
Let us determine a set of linearly dependent $q$-ary
 rows in this submatrix.\\
\textbf{2.1. } All elements of the first row of the submatrix
$h_{2}$ are elements of the field $GF(t)$ .
Indeed,
\begin{center}
$ \left(\frac{1}{(\alpha _{i}^{t+1}+1)^{2}}
\right)^{t}=\frac{1}{(\alpha _{i}^{t+1}+1)^{2}} $
\end{center}
for any $i=1,\ldots,n-1$ .\\
\textbf{2.2.} The row
\begin{center}
$ \left[
\begin{array}{ccccc}
 \frac{\alpha _{1}^{t}}{(\alpha _{1}^{t+1}+1)^{2}} &
  \frac{\alpha _{2}^{t}}{(\alpha _{2}^{t+1}+1)^{2}}&\ldots &
  \frac{\alpha _{n-1}^{t}}{(\alpha _{n-1}^{t+1}+1)^{2}}& 0
\end{array}
\right] $
\end{center}
is obtained by the element-by-element raising of the row of the
submatrix
\begin{center}
$ \left[
 \begin{array}{ccccc}
\frac{\alpha _{1}}{(\alpha _{1}^{t+1}+1)^{2}} &
 \frac{\alpha _{2}}{(\alpha _{2}^{t+1}+1)^{2}}&\ldots &
  \frac{\alpha _{n-1}}{(\alpha _{n-1}^{t+1}+1)^{2}}& 0
\end{array}
\right]
 $
\end{center}
to the power $t$.\\
\textbf{2.3. } The row
\begin{center}
$ \left[
\begin{array}{ccccc}
 \frac{\alpha _{1}^{t-1}}{(\alpha _{1}^{t+1}+1)^{2}} &
 \frac{\alpha _{2}^{t-1}}{(\alpha _{2}^{t+1}+1)^{2}}&\ldots &
 \frac{\alpha _{n-1}^{t-1}}{(\alpha _{n-1}^{t+1}+1)^{2}}& 0
\end{array}
\right]
 $
\end{center}
is obtained as a linear combination of the corresponding row of
the submatrix $h_{1}$
\begin{center}
$ \left[
\begin{array}{ccccc}
 \frac{\alpha _{1}^{t-1}}{\alpha _{1}^{t+1}+1} &
 \frac{\alpha _{2}^{t-1}}{\alpha _{2}^{t+1}+1}&\ldots &
 \frac{\alpha _{n-1}^{t-1}}{\alpha _{n-1}^{t+1}+1}& 0
\end{array}
\right]
 $
\end{center}
 and  the $t$-th power of the corresponding row of submatrix $h_{2}$
\begin{center}
$ \left[
\begin{array}{ccccc}
 \frac{\alpha _{1}^{2}}{(\alpha _{1}^{t+1}+1)^{2}} &
 \frac{\alpha _{2}^{2}}{(\alpha _{2}^{t+1}+1)^{2}}&\ldots &
 \frac{\alpha _{n-1}^{2}}{(\alpha _{n-1}^{t+1}+1)^{2}}& 0
\end{array}
\right]
 $.
\end{center}
It means that we can write the following relation for any
element of these rows:
\begin{center}
$ \frac{\alpha _{i}^{t-1}}{(\alpha _{i}^{t+1}+1)^{2}}=
\frac{\alpha _{i}^{t-1}}{(\alpha
_{i}^{t+1}+1)}-\left(\frac{\alpha _{i}^{2}}{(\alpha
_{i}^{t+1}+1)^{2}}\right)^{t} $
\end{center}
for all $i=1,\ldots,n-1$.

 Thus, the number of $q$-ary rows
equal to $\delta_{2} =l+2l+2l$ can be deleted from $h_{2}$ as
they are linearly dependent on the other rows of this and
preceding submatrix.
\begin{remark} ~\label{note2_h1}
Note that we used the row:
\begin{equation*}
\left[
\begin{array}{ccccc}
 \frac{\alpha _{1}^{t-1}}{\alpha _{1}^{t+1}+1} &
 \frac{\alpha _{2}^{t-1}}{\alpha _{2}^{t+1}+1}&\ldots &
  \frac{\alpha _{n-1}^{t-1}}{\alpha _{n-1}^{t+1}+1}& 0
\end{array}
\right]
\end{equation*}
of the submatrix $h_{1}$ when representing the row of the
submatrix $h_{2}$.
\end{remark}
\textbf{3.} Consider the submatrix $h_{3}$:
\begin{equation}
h_{3}=\left[ \begin{array}{ccccc}
\frac{1}{(\alpha _{1}^{t+1}+1)^{3}} &
\frac{1}{(\alpha _{2}^{t+1}+1)^{3}}& \ldots &
 \frac{1}{(\alpha _{n-1}^{t+1}+1)^{3}}& 1 \\
\frac{\alpha _{1}}{(\alpha _{1}^{t+1}+1)^{3}} &
\frac{\alpha _{2}}{(\alpha _{2}^{t+1}+1)^{3}}&\ldots &
\frac{\alpha _{n-1}}{(\alpha _{n-1}^{t+1}+1)^{3}}& 0 \\
\ldots & \ldots & \ldots & \ldots & \ldots\\
\frac{\alpha _{1}^{t}}{(\alpha _{1}^{t+1}+1)^{3}} &
 \frac{\alpha _{2}^{t}}{(\alpha _{2}^{t+1}+1)^{3}}&\ldots &
  \frac{\alpha _{n-1}^{t}}{(\alpha _{n-1}^{t+1}+1)^{3}}& 0 \\
\end{array}
\right], \label{parity-check_ matrix 3 G6_v1}
\end{equation}
where $\alpha_{i}\in L_{6} $ and $\alpha_{n}=0$.\\
Let us determine a set of linearly dependent $q$-ary
rows in this submatrix.\\
\textbf{3.1.} All elements of the first row of the submatrix
$h_{3}$ are elements of the field $GF(t)$ .
Indeed,
\begin{center}
$ \left(\frac{1}{(\alpha _{i}^{t+1}+1)^{3}}
\right)^{t}=\frac{1}{(\alpha _{i}^{t+1}+1)^{3}} \text{ for any
} i=1,\ldots,n-1. $
\end{center}
\textbf{3.2.} The row
\begin{center}
$ \left[ \begin{array}{ccccc}
 \frac{\alpha _{1}^{t}}{(\alpha _{1}^{t+1}+1)^{3}} &
 \frac{\alpha _{2}^{t}}{(\alpha _{2}^{t+1}+1)^{3}}&\ldots &
 \frac{\alpha _{n-1}^{t}}{(\alpha _{n-1}^{t+1}+1)^{3}}& 0 \\
\end{array}
\right] $
\end{center}
is obtained by element-by-element raising of the row of the
submatrix
\begin{center}
$ \left[
\begin{array}{ccccc}
 \frac{\alpha _{1}}{(\alpha _{1}^{t+1}+1)^{3}} &
 \frac{\alpha _{2}}{(\alpha _{2}^{t+1}+1)^{3}}&\ldots &
  \frac{\alpha _{n-1}}{(\alpha _{n-1}^{t+1}+1)^{3}}& 0 \\
\end{array}
\right] $
\end{center}
to the $t$-th power.\\
\textbf{3.3.}  The row
\begin{center}
$ \left[
\begin{array}{ccccc}
 \frac{\alpha _{1}^{t-1}}{(\alpha _{1}^{t+1}+1)^{3}} &
 \frac{\alpha _{2}^{t-1}}{(\alpha _{2}^{t+1}+1)^{3}}&\ldots &
 \frac{\alpha _{n-1}^{t-1}}{(\alpha _{n-1}^{t+1}+1)^{3}}& 0 \\
\end{array}
\right] $
\end{center}
is obtained as a linear combination of the corresponding
submatrix $h_{2}$ row \begin{center} $ \left[
\begin{array}{ccccc}
 \frac{\alpha _{1}^{t-1}}{(\alpha _{1}^{t+1}+1)^{2}} &
 \frac{\alpha _{2}^{t-1}}{(\alpha _{2}^{t+1}+1)^{2}}&\ldots &
 \frac{\alpha _{n-1}^{t-1}}{(\alpha _{n-1}^{t+1}+1)^{2}}& 0 \\
\end{array}
\right] $
\end{center}
and the $t$-th power of the corresponding row of the submatrix
$h_{3}$
\begin{center}
$ \left[
\begin{array}{ccccc}
 \frac{\alpha _{1}^{2}}{(\alpha _{1}^{t+1}+1)^{3}} &
 \frac{\alpha _{2}^{2}}{(\alpha _{2}^{t+1}+1)^{3}}&\ldots &
  \frac{\alpha _{n-1}^{2}}{(\alpha _{n-1}^{t+1}+1)^{3}}& 0 \\
\end{array}
\right] $ .
\end{center}
It means that the following relation:
\begin{center}
$ \frac{\alpha _{i}^{t-1}}{(\alpha _{i}^{t+1}+1)^{3}}=
\frac{\alpha _{i}^{t-1}}{(\alpha
_{i}^{t+1}+1)^{2}}-\left(\frac{\alpha _{i}^{2}}{(\alpha
_{i}^{t+1}+1)^{3}}\right)^{t} $
\end{center}
can be written for any element of these rows for all $i=1,\ldots,n-1$.\\
\textbf{3.4.} The row
\begin{center}
 $ \left[
 \begin{array}{ccccc}
 \frac{\alpha _{1}^{t-2}}{(\alpha _{1}^{t+1}+1)^{3}} &
 \frac{\alpha _{2}^{t-2}}{(\alpha _{2}^{t+1}+1)^{3}}&\ldots &
  \frac{\alpha _{n-1}^{t-2}}{(\alpha _{n-1}^{t+1}+1)^{3}}& 0 \\
\end{array}
\right]
 $
\end{center}
is obtained as a linear combination of the corresponding row of
the submatrix  $h_{1}$
\begin{center}
$ \left[
\begin{array}{ccccc}
 \frac{\alpha _{1}^{t-2}}{\alpha _{1}^{t+1}+1} & \frac{\alpha _{2}^{t-2}}{\alpha _{2}^{t+1}+1}&
 \ldots & \frac{\alpha _{n-1}^{t-2}}{\alpha _{n-1}^{t+1}+1}& 0 \\
\end{array}
\right]
 $,
\end{center}
the row of the submatrix $h_{2}$
\begin{center}
$ \left[
\begin{array}{ccccc}
\frac{\alpha _{1}^{t-2}}{(\alpha _{1}^{t+1}+1)^{2}} &
\frac{\alpha _{2}^{t-2}}{(\alpha _{2}^{t+1}+1)^{2}}&
 \ldots & \frac{\alpha _{n-1}^{t-2}}{(\alpha _{n-1}^{t+1}+1)^{2}}& 0 \\
\end{array}
\right]
 $
\end{center}
and the $t$-th power of the row
\begin{center}
$ \left[
\begin{array}{ccccc}
\frac{\alpha _{1}^{3}}{(\alpha _{1}^{t+1}+1)^{3}} &
\frac{\alpha _{2}^{3}}{(\alpha _{2}^{t+1}+1)^{3}}&
 \ldots & \frac{\alpha _{n-1}^{3}}{(\alpha _{n-1}^{t+1}+1)^{3}}& 0 \\
\end{array}
\right]
 $
\end{center}
of the submatrix $h_{3}$.
 It means that we can write the following relations:
\begin{center}
$ \frac{\alpha _{i}^{t-2}}{(\alpha _{i}^{t+1}+1)^{3}}=
\frac{\alpha _{i}^{t-2}}{(\alpha
_{i}^{t+1}+1)^{2}}-\frac{\alpha _{i}^{2t-1}}{(\alpha
_{i}^{t+1}+1)^{3}}$,\\
$ \frac{\alpha _{i}^{2t-1}}{(\alpha _{i}^{t+1}+1)^{3}}=
\frac{\alpha _{i}^{2t-1}}{(\alpha
_{i}^{t+1}+1)^{2}}-\left(\frac{\alpha _{i}^{3}}{(\alpha
_{i}^{t+1}+1)^{3}}\right)^{t}$,\\
$\frac{\alpha _{i}^{2t-1}}{(\alpha _{i}^{t+1}+1)^{2}}=
\frac{\alpha _{i}^{t-2}}{(\alpha _{i}^{t+1}+1)}-\frac{\alpha
_{i}^{t-2}}{(\alpha _{i}^{t+1}+1)^{2}}$,\\
\end{center}
for any element of this row for all $i=1,\ldots,n-1$.

 Thus, the
number of  the $q$-ary row equal to $\delta_{3} =l+2l+2l+2l$
can be deleted from the matrix $h_{3}$ as they are linearly
dependent on the other rows of this and preceding submatrices.
\begin{remark} \label{note3_h1}
Note that the row of the submatrix $h_{1}$
\begin{equation*}
 \left[
\begin{array}{ccccc}
 \frac{\alpha _{1}^{t-2}}{\alpha _{1}^{t+1}+1} & \frac{\alpha _{2}^{t-2}}{\alpha _{2}^{t+1}+1}&
 \ldots & \frac{\alpha _{n-1}^{t-2}}{\alpha _{n-1}^{t+1}+1}& 0 \\
\end{array}
\right]
\end{equation*}
was used for representing the row of the submatrix $h_{3}$.
\end{remark}

It is clear that the similar situation takes place for any submatrix.\\
\textbf{j. }($j \leq q-1$) Consider a submatrix $h_{j}$:
\begin{equation}
h_{j}=\left[ \begin{array}{ccccc}
\frac{1}{(\alpha _{1}^{t+1}+1)^{j}} & \frac{1}{(\alpha _{2}^{t+1}+1)^{j}}& \ldots & \frac{1}{(\alpha _{n-1}^{t+1}+1)^{j}}& 1 \\
\frac{\alpha _{1}}{(\alpha _{1}^{t+1}+1)^{j}} & \frac{\alpha _{2}}{(\alpha _{2}^{t+1}+1)^{j}}&\ldots & \frac{\alpha _{n-1}}{(\alpha _{n-1}^{t+1}+1)^{j}}& 0 \\
\ldots & \ldots & \ldots & \ldots & \ldots\\
\frac{\alpha _{1}^{t}}{(\alpha _{1}^{t+1}+1)^{j}} & \frac{\alpha _{2}^{t}}{(\alpha _{2}^{t+1}+1)^{j}}&\ldots & \frac{\alpha _{n-1}^{t}}{(\alpha _{n-1}^{t+1}+1)^{j}}& 0 \\
\end{array}
\right], \label{parity-check _matrix i G6_v1}
\end{equation}
where $\alpha_{i}\in L_{6} $ and $\alpha_{n}=0$.\\
Let us determine a set of linearly dependent $q$-ary
rows of this submatrix.\\
\textbf{j.1.} All elements of the first row of the submatrix
$h_{j}$ are elements of the field $GF(t)$ . Indeed,
\begin{center}
$ \left(\frac{1}{(\alpha _{i}^{t+1}+1)^{j}}
\right)^{t}=\frac{1}{(\alpha _{i}^{t+1}+1)^{j}} \text{ for any
} i=1,\ldots,n-1. $
\end{center}
\textbf{j.2.} The row
\begin{center}
$ \left[
 \begin{array}{ccccc}
 \frac{\alpha _{1}^{t}}{(\alpha _{1}^{t+1}+1)^{j}} &
  \frac{\alpha _{2}^{t}}{(\alpha _{2}^{t+1}+1)^{j}}&\ldots &
  \frac{\alpha _{n-1}^{t}}{(\alpha _{n-1}^{t+1}+1)^{j}}& 0 \\
\end{array}
\right]
 $
\end{center}
is obtained by element-by-element raising of the row
\begin{center}
$ \left[
\begin{array}{ccccc}
 \frac{\alpha _{1}}{(\alpha _{1}^{t+1}+1)^{j}} &
 \frac{\alpha _{2}}{(\alpha _{2}^{t+1}+1)^{j}}&\ldots &
 \frac{\alpha _{n-1}}{(\alpha _{n-1}^{t+1}+1)^{j}}& 0 \\
\end{array}
\right]
 $
\end{center}
of the submatrix to the $t$-th power.\\
\textbf{j.3.}  The row
\begin{center}
$ \left[
 \begin{array}{ccccc}
 \frac{\alpha _{1}^{t-1}}{(\alpha _{1}^{t+1}+1)^{j}} &
  \frac{\alpha _{2}^{t-1}}{(\alpha _{2}^{t+1}+1)^{j}}&\ldots &
  \frac{\alpha _{n-1}^{t-1}}{(\alpha _{n-1}^{t+1}+1)^{j}}& 0 \\
\end{array}
\right]
 $
\end{center}
is obtained as a linear combination of corresponding row of the
submatrix $h_{j-1}$  and the  $t$-th power of the corresponding
row of the submatrix $h_{j}$.
 Thus, the relation
\begin{center}
$ \frac{\alpha _{i}^{t-1}}{(\alpha _{i}^{t+1}+1)^{j}}=
\frac{\alpha _{i}^{t-1}}{(\alpha
_{i}^{t+1}+1)^{j-1}}-\left(\frac{\alpha _{i}^{2}}{(\alpha
_{i}^{t+1}+1)^{j}}\right)^{t} $
\end{center}
can be written for any element of these rows for all $i=1,\ldots,n-1$.\\
And so on.

Finally, at the last step we have the following relations for rows  of the $j$-th submatrix:\\
\textbf{j.j+1.} The row
\begin{center}
 $
 \left[
 \begin{array}{ccccc}
\frac{\alpha _{1}^{t-(j-1)}}{(\alpha _{1}^{t+1}+1)^{j}} &
 \frac{\alpha _{2}^{t-(j-1)}}{(\alpha _{2}^{t+1}+1)^{j}}&
 \ldots & \frac{\alpha _{n-1}^{t-(j-1)}}{(\alpha _{n-1}^{t+1}+1)^{j}}& 0 \\
\end{array}
\right]
 $
\end{center}
is obtained as linear combination of preceding rows of the
submatrices $h_{j}$ and $h_{j-1}, \ldots, h_{2}, h_{1}$. In
other words, we can write following relations for any elements
of this row:
\begin{center}
$ \frac{\alpha _{i}^{t-(j-1)}}{(\alpha _{i}^{t+1}+1)^{j}}=
\frac{\alpha _{i}^{t-(j-1)}}{(\alpha
_{i}^{t+1}+1)^{j-1}}-\frac{\alpha _{i}^{2t-(j-2)}}{(\alpha
_{i}^{t+1}+1)^{j}}$,\\
$ \frac{\alpha _{i}^{2t-(j-2)}}{(\alpha _{i}^{t+1}+1)^{j}}=
\frac{\alpha _{i}^{2t-(j-2)}}{(\alpha
_{i}^{t+1}+1)^{j-1}}-\frac{\alpha _{i}^{3t-(j-3)}}{(\alpha
_{i}^{t+1}+1)^{j}}$,\\
$ \vdots$ \\
$ \frac{\alpha _{i}^{(j-1)t-1}}{(\alpha
_{i}^{t+1}+1)^{j}}=\frac{\alpha _{i}^{(j-1)t-1}}{(\alpha
_{i}^{t+1}+1)^{j-1}}-\left(\frac{\alpha _{i}^{j}}{(\alpha
_{i}^{t+1}+1)^{j}} \right)^{t}$,\\
\end{center}
for all $i=1,\ldots,n-1$.\\
Here, the row
\begin{center}
 $
 \left[
 \begin{array}{ccccc}
 \frac{\alpha _{1}^{t-(j-1)}}{(\alpha _{1}^{t+1}+1)^{j-1}} &
  \frac{\alpha _{2}^{t-(j-1)}}{(\alpha _{2}^{t+1}+1)^{j-1}}&
 \ldots & \frac{\alpha _{n-1}^{t-(j-1)}}{(\alpha _{n-1}^{t+1}+1)^{j-1}}& 0 \\
\end{array}
\right]
 $
\end{center}
belongs to the submatrix $h_{j-1}$, whereas the row
\begin{center}
 $
 \left[
 \begin{array}{ccccc}
 \frac{\alpha _{1}^{j}}{(\alpha _{1}^{t+1}+1)^{j}} &
 \frac{\alpha _{2}^{j}}{(\alpha _{2}^{t+1}+1)^{j}}&
 \ldots & \frac{\alpha _{n-1}^{j}}{(\alpha _{n-1}^{t+1}+1)^{j}}& 0 \\
\end{array}
\right]
 $
\end{center}
is a previous one in the submatrix $h_{j}$ because $t-(j-1)>j$.
Let us consider now $j-2$ rows with elements
\begin{equation*}
\begin{array}{c}
 \left\{\frac{\alpha _{i}^{2t-(j-2)}}{(\alpha
_{i}^{t+1}+1)^{j-1}}\right\},\; \left\{ \frac{\alpha
_{i}^{3t-(j-3)}}{(\alpha _{i}^{t+1}+1)^{j-1}}\right\},\;\ldots
,\;
 \left\{\frac{\alpha _{i}^{(j-1)t-1}}{(\alpha
_{i}^{t+1}+1)^{j-1}}\right\},\\
 i=1,\ldots,n.
\end{array}
\end{equation*}
The following relations can be written for the components of these
rows:
\begin{center}
$ \frac{\alpha _{i}^{t-(j-1)}}{(\alpha _{i}^{t+1}+1)^{j-1}}=
\frac{\alpha _{i}^{t-(j-1)}}{(\alpha
_{i}^{t+1}+1)^{j-2}}-\frac{\alpha _{i}^{2t-(j-2)}}{(\alpha
_{i}^{t+1}+1)^{j-1}}$ ,\\
$ \frac{\alpha _{i}^{2t-(j-2)}}{(\alpha _{i}^{t+1}+1)^{j-1}}=
\frac{\alpha _{i}^{2t-(j-2)}}{(\alpha
_{i}^{t+1}+1)^{j-2}}-\frac{\alpha _{i}^{3t-(j-3)}}{(\alpha
_{i}^{t+1}+1)^{j-1}}$ ,\\
$ \vdots $ \\
$ \frac{\alpha _{i}^{(j-2)t-2}}{(\alpha
_{i}^{t+1}+1)^{j-1}}=\frac{\alpha _{i}^{(j-2)t-2}}{(\alpha
_{i}^{t+1}+1)^{j-2}}-\frac{\alpha _{i}^{(j-1)t-1}}{(\alpha
_{i}^{t+1}+1)^{j-1}}$ ,\\
\end{center}
for all $i=1,\ldots,n , \; \alpha_{n}=0$.\\
Here the rows
\begin{center}
 $
 \left[
 \begin{array}{ccccc}
 (\frac{\alpha _{1}^{t-(j-1)}}{(\alpha _{1}^{t+1}+1)^{j-1}} &
 \frac{\alpha _{2}^{t-(j-1)}}{(\alpha _{2}^{t+1}+1)^{j-1}}&
 \ldots & \frac{\alpha _{n-1}^{t-(j-1)}}{(\alpha _{n-1}^{t+1}+1)^{j-1}}& 0 \\
\end{array}
\right]
 $
\end{center}
and
\begin{center}
 $
 \left[
 \begin{array}{ccccc}
 \frac{\alpha _{1}^{t-(j-1)}}{(\alpha _{1}^{t+1}+1)^{j-2}} &
  \frac{\alpha _{2}^{t-(j-1)}}{(\alpha _{2}^{t+1}+1)^{j-2}}&
 \ldots & \frac{\alpha _{n-1}^{t-(j-1)}}{(\alpha _{n-1}^{t+1}+1)^{j-2}}& 0 \\
\end{array}
\right]
 $
\end{center}
belong to submatrices $h_{j-1}$ and $h_{j-2}$, respectively.
Similar relations can be written for auxiliary $j-3$ rows with
elements
\begin{equation*}
\begin{array}{c}
\left\{ \frac{\alpha _{i}^{2t-(j-2)}}{(\alpha _{i}^{t+1}+1)^{j-2}}\right\},
\left\{ \frac{\alpha _{i}^{3t-(j-3)}}{(\alpha _{i}^{t+1}+1)^{j-2}}\right\},
\ldots,
 \left\{\frac{\alpha _{i}^{(j-2)t-2}}{(\alpha_{i}^{t+1}+1)^{j-2}}\right\},\\
i=1,\ldots,n
 \end{array}
\end{equation*}
by using the rows in submatrices $h_{j-1}$ and $h_{j-2}$ and
auxiliary $j-4$ rows with elements
\begin{equation*}
\begin{array}{c}
\left\{\frac{\alpha _{i}^{2t-(j-2)}}{(\alpha _{i}^{t+1}+1)^{j-3}}\right\},
\left\{\frac{\alpha _{i}^{3t-(j-3)}}{(\alpha _{i}^{t+1}+1)^{j-3}}\right\},
\ldots,
\left\{\frac{\alpha _{i}^{(j-3)t-3}}{(\alpha_{i}^{t+1}+1)^{j-3}}\right\},\\
i=1,\ldots,n.
\end{array}
\end{equation*}
Thus, the number of relations decreases by one with every step,
so that we obtain for the last step:
\begin{center}
$ \frac{\alpha _{i}^{2t-(j-2)}}{(\alpha _{i}^{t+1}+1)^{2}}=
\frac{\alpha _{i}^{t-(j-1)}}{(\alpha
_{i}^{t+1}+1)}-\frac{\alpha _{i}^{t-(j-1)}}{(\alpha
_{i}^{t+1}+1)^{2}}$\\
\end{center}
for all $i=1,\ldots,n$.\\
Here the rows
\begin{center}
 $
 \left[
 \begin{array}{ccccc}
 \frac{\alpha _{1}^{t-(j-1)}}{\alpha _{1}^{t+1}+1} &
  \frac{\alpha _{2}^{t-(j-1)}}{\alpha _{2}^{t+1}+1}&
 \ldots & \frac{\alpha _{n-1}^{t-(j-1)}}{\alpha _{n-1}^{t+1}+1}& 0 \\
\end{array}
\right]
 $
\end{center}
and
\begin{center}
 $
 \left[
 \begin{array}{ccccc}
 \frac{\alpha _{1}^{t-(j-1)}}{(\alpha _{1}^{t+1}+1)^{2}} &
 \frac{\alpha _{2}^{t-(j-1)}}{(\alpha _{2}^{t+1}+1)^{2}}&
 \ldots & \frac{\alpha _{n-1}^{t-(j-1)}}{(\alpha _{n-1}^{t+1}+1)^{2}}& 0 \\
\end{array}
\right]
 $
\end{center}
belong to the submatrices $h_{1}$ and $h_{2}$, respectively.

Hence, the number of  $q$-ary rows equal to $\delta_{j} =l+2lj$
can be deleted from the submatrix $h_{j}$ because they are
linearly dependent on other rows of this and previous
submatrices.
\begin{remark} \label{notej_h1}
Note that we used the row
\begin{equation*}
 \left[
 \begin{array}{ccccc}
 \frac{\alpha _{1}^{t-(j-1)}}{\alpha _{1}^{t+1}+1} &
  \frac{\alpha _{2}^{t-(j-1)}}{\alpha _{2}^{t+1}+1}&
 \ldots & \frac{\alpha _{n-1}^{t-(j-1)}}{\alpha _{n-1}^{t+1}+1}& 0 \\
\end{array}
\right]
\end{equation*}
of the submatrix $h_{1}$ when representing the row of the
submatrix $h_{j}$, $j<q$.
\end{remark}
The total number of $q$-ary rows in  submatrices $h_{1}, h_{2},
\ldots,h_{j}$,  $j<q$  that are linearly dependent and satisfy
the above presented relations is equal to
 $\Delta_{j}$:
\begin{center}
$ \Delta_{j}= \delta_{1}+\delta_{2}+\ldots+\delta_{j}=
lj+2l(1+2+\ldots+j)= lj+2l\frac{j(j+1)}{2} $,
\end{center}
\begin{equation}
\Delta_{j}= 2l\frac{j(j+3)}{2} .
\end{equation}
So, the estimation for the redundancy of the
cumulative-separable $(L,G)$-code with $G(x)=G_{6}^{(j)}(x)
=\left(x^{t+1}+1\right)^{j}$, $1<j\leq q-2$  and the set
$L_{6}=\left\{GF(t^{2}) \backslash \left\{\alpha :
G_{6}(\alpha)=0\right\} \right\}$ is given by
 $r^{(j)}_{6}$:
\begin{equation}
r^{(j)}_{6}\leq 2l\left(j(t+1)-\frac{j(j+3)}{2}\right).
\end{equation}
Consequently, the estimation on the dimension of this code is
\begin{equation}
k_{6}^{(j)} \geq t^{2}-t-1-2lj\left(t+1-\frac{j+3}{2}\right) .
\end{equation}
Of special interest is the case $j=q-1,q$ , i.e., the case when
the submatrix $h_{q-1}$ is presented in  the parity check
matrix $H_{6}$ of the code.

 Any row
 \begin{center}
 $
 \left[
 \begin{array}{ccccc}
 \frac{\alpha _{1}^{j\frac{t}{q}+1}}{\alpha _{1}^{t+1}+1} &
 \frac{\alpha _{2}^{j\frac{t}{q}+1}}{\alpha _{2}^{t+1}+1}&
 \ldots & \frac{\alpha _{n-1}^{j\frac{t}{q}+1}}{\alpha _{n-1}^{t+1}+1}& 0 \\
\end{array}
\right] , \; j=1,\ldots,q-1
 $
\end{center}
from $h_{1}$ can be obtained as a linear combination of the
rows of submatrices $h_{q-1} , h_{q-2}, \ldots, h_{1}$.

In other words, the estimation of the redundancy of this code
can be improved by the value $\theta_{q-1}=(q-1)2l$ in case
there is $h_{q-1}$  in the cumulative-separable parity check
matrix.
\begin{remark}
According to Remarks
\ref{note2_h1},\ref{note3_h1},\ref{notej_h1}, we did not use
the above considered rows of the submatrix $h_{1}$  for the
analysis of the linear dependence of rows in the parity check
matrix $H_{6}$ of the cumulative-separable
$\Gamma(L_{6},G_{6}^{(q)})$-code.
\end{remark}
 First of all, let us prove the following statement.
\begin{lemma}
All rows of the submatrix $h_{1}$
\begin{equation*}
  \left[
 \begin{array}{ccccc}
 \frac{\alpha _{1}^{f\frac{t}{q}+1}}{\alpha _{1}^{t+1}+1} &
  \frac{\alpha _{2}^{f\frac{t}{q}+1}}{\alpha _{2}^{t+1}+1}&
 \ldots & \frac{\alpha _{n-1}^{f\frac{t}{q}+1}}{\alpha _{n-1}^{t+1}+1}& 0 \\
\end{array}
\right]
\end{equation*}
for $f=1,\ldots,q-1$ can be obtained as alinear combination of
the corresponding row of the submatrix $h_{1}$
\begin{equation*}
 \left[
 \begin{array}{ccccc}
 \frac{\alpha _{1}^{t-f\frac{t}{q}}}{\alpha _{1}^{t+1}+1} &
 \frac{\alpha _{2}^{t-f\frac{t}{q}}}{\alpha _{2}^{t+1}+1}&
 \ldots & \frac{\alpha _{n-1}^{t-f\frac{t}{q}}}{\alpha _{n-1}^{t+1}+1}& 0 \\
\end{array}
\right] ,
\end{equation*}
and auxiliary rows
\begin{equation*}
 \left[
 \begin{array}{ccccc}
 \frac{\alpha _{1}^{(f-j)t + q-j}}{(\alpha _{1}^{t+1}+1)^{q-1}} &
 \frac{\alpha _{2}^{(f-j)t + q-j}}{(\alpha _{2}^{t+1}+1)^{q-1}}&
 \ldots & \frac{\alpha _{n-1}^{(f-j)t + q-j}}{(\alpha _{n-1}^{t+1}+1)^{q-1}}& 0 \\
\end{array}
\right] ,
\end{equation*}
where $j=1,\ldots,f$.

 \begin{proof} \\
 It is easy to observe that the row
\begin{equation*}
 \left[
 \begin{array}{ccccc}
 \frac{\alpha _{1}^{ft+q}}{(\alpha _{1}^{t+1}+1)^{q}} &
 \frac{\alpha _{2}^{ft+q}}{(\alpha _{2}^{t+1}+1)^{q}}&
 \ldots & \frac{\alpha _{n-1}^{ft+q}}{(\alpha _{n-1}^{t+1}+1)^{q}}& 0 \\
\end{array}
\right]
\end{equation*}
 results from the row
 \begin{equation*}
 \left[
 \begin{array}{ccccc}
 \frac{\alpha _{1}^{f\frac{t}{q}+1}}{\alpha _{1}^{t+1}+1} &
 \frac{\alpha _{2}^{f\frac{t}{q}+1}}{\alpha _{2}^{t+1}+1}&
 \ldots & \frac{\alpha _{n-1}^{f\frac{t}{q}+1}}{\alpha _{n-1}^{t+1}+1}& 0
\end{array}
\right].
\end{equation*}
by raising of all its elements to the power $q$.

Now let us write a system of recurrent equations:
\begin{equation*}
\begin{array}{c}
\frac{\alpha _{i}^{ft+q}}{(\alpha _{i}^{t+1}+1)^{q}}=
\frac{\alpha _{i}^{(f-1)t+q-1}}{(\alpha
_{i}^{t+1}+1)^{q-1}}-\frac{\alpha _{i}^{(f-1)t+q-1}}{(\alpha
_{i}^{t+1}+1)^{q}},\\
 \frac{\alpha _{i}^{(f-1)t+q-1}}{(\alpha _{i}^{t+1}+1)^{q}}=
\frac{\alpha _{i}^{(f-2)t+q-2}}{(\alpha
_{i}^{t+1}+1)^{q-1}}-\frac{\alpha _{i}^{(f-2)t+q-2}}{(\alpha
_{i}^{t+1}+1)^{q}},\\
 \frac{\alpha _{i}^{(f-2)t+q-2}}{(\alpha _{i}^{t+1}+1)^{q}}=
\frac{\alpha _{i}^{(f-3)t+q-3}}{(\alpha
_{i}^{t+1}+1)^{q-1}}-\frac{\alpha _{i}^{(f-3)t+q-3}}{(\alpha
_{i}^{t+1}+1)^{q}},\\
\vdots\\
 \frac{\alpha _{i}^{(f-r)t+q-r}}{(\alpha _{i}^{t+1}+1)^{q}}=
\frac{\alpha _{i}^{(f-r-1)t+q-r-1}}{(\alpha
_{i}^{t+1}+1)^{q-1}}-\frac{\alpha
_{i}^{(f-r-1)t+q-r-1}}{(\alpha_{i}^{t+1}+1)^{q}},\\
\vdots\\
\frac{\alpha_{i}^{t+q-(f-1)}}{(\alpha_{i}^{t+1}+1)^{q}}=
\frac{\alpha _{i}^{q-f}}{(\alpha
_{i}^{t+1}+1)^{q-1}}-\frac{\alpha _{i}^{q-f}}{(\alpha
_{i}^{t+1}+1)^{q}}.
\end{array}
\end{equation*}

Obviously, the second term in the right part of the last
equation can be obtained as the $qt$-th power of the element
$\frac{\alpha _{i}^{t-f\frac{t}{q}}}{\alpha _{i}^{t+1}+1}$:
\begin{center}
$ \left(\frac{\alpha _{i}^{t-f\frac{t}{q}}}{\alpha
_{i}^{t+1}+1}\right)^{qt}=\frac{\alpha _{i}^{q-f}}{(\alpha
_{i}^{t+1}+1)^{q}} ,
\begin{array} {l}  \text{ where } t-f\frac{t}{q}<t-1\\
\text{ for all } f=1,\ldots,q-1.
\end{array}
$
\end{center}

\end{proof}
\end{lemma}
Now write a representation of the above considered auxiliary
rows in terms of other auxiliary rows which elements have
smaller degree $j$ of denominator $\left(\alpha
_{i}^{t+1}+1\right)^{j}$.

\begin{lemma}
All auxiliary rows with components  $\frac{\alpha
_{i}^{(f-1)t+q-1}}{(\alpha _{i}^{t+1}+1)^{q-1}} $,
$f=2,\ldots,q-1$  can be obtained as a linear combination of
corresponding auxiliary rows with the components $\frac{\alpha
_{i}^{(v-2)t+q-2}}{(\alpha _{i}^{t+1}+1)^{q-2}}$
 , $v=3,\ldots,q-1$ and rows from submatrices $h_{q-1},h_{q-2}$:
\begin{equation*}
 \left[
 \begin{array}{ccccc}
\frac{\alpha _{1}^{q-f}}{(\alpha _{1}^{t+1}+1)^{q-1}} &
 \frac{\alpha _{2}^{q-f}}{(\alpha _{2}^{t+1}+1)^{q-1}}&
\ldots & \frac{\alpha _{n-1}^{q-f}}{(\alpha _{n-1}^{t+1}+1)^{q-1}}& 0 \\
\end{array}
\right] ,
\end{equation*}
\begin{equation*}
 \left[
 \begin{array}{ccccc}
 \frac{\alpha _{1}^{q-f}}{(\alpha _{1}^{t+1}+1)^{q-2}} &
  \frac{\alpha _{2}^{q-f}}{(\alpha _{2}^{t+1}+1)^{q-2}}&
 \ldots & \frac{\alpha _{n-1}^{q-f}}{(\alpha _{n-1}^{t+1}+1)^{q-2}}& 0 \\
\end{array}
\right].
\end{equation*}
\begin{proof}
Similarly to the previous Lemma, we can write a system of
recurrent equations:
\begin{equation*}
\begin{array}{c}
 \frac{\alpha _{i}^{(f-1)t+q-1}}{(\alpha _{i}^{t+1}+1)^{q-1}}=
\frac{\alpha _{i}^{(f-2)t+q-2}}{(\alpha
_{i}^{t+1}+1)^{q-2}}-\frac{\alpha _{i}^{(f-2)t+q-2}}{(\alpha
_{i}^{t+1}+1)^{q-1}},\\
\frac{\alpha _{i}^{(f-2)t+q-2}}{(\alpha _{i}^{t+1}+1)^{q-1}}=
\frac{\alpha _{i}^{(f-3)t+q-3}}{(\alpha
_{i}^{t+1}+1)^{q-2}}-\frac{\alpha _{i}^{(f-3)t+q-3}}{(\alpha
_{i}^{t+1}+1)^{q-1}},\\
 \frac{\alpha _{i}^{(f-3)t+q-3}}{(\alpha _{i}^{t+1}+1)^{q-1}}=
\frac{\alpha _{i}^{(f-4)t+q-4}}{(\alpha
_{i}^{t+1}+1)^{q-2}}-\frac{\alpha _{i}^{(f-4)t+q-4}}{(\alpha
_{i}^{t+1}+1)^{q-1}},\\
\vdots\\
\frac{\alpha _{i}^{(f-r)t+q-r}}{(\alpha _{i}^{t+1}+1)^{q-1}}=
\frac{\alpha _{i}^{(f-r-1)t+q-r-1}}{(\alpha
_{i}^{t+1}+1)^{q-2}}-\frac{\alpha
_{i}^{(f-r-1)t+q-r-1}}{(\alpha
_{i}^{t+1}+1)^{q-1}},\\
\vdots\\
\frac{\alpha _{i}^{t+q-(f-1)}}{(\alpha _{i}^{t+1}+1)^{q-1}}=
\frac{\alpha _{i}^{q-f}}{(\alpha
_{i}^{t+1}+1)^{q-2}}-\frac{\alpha _{i}^{q-f}}{(\alpha
_{i}^{t+1}+1)^{q-1}}.
\end{array}
\end{equation*}
It is clear that the elements in the right part of the last
equation are components of the corresponding rows of submatrices
$h_{q-1}$ and $h_{q-2}$.
\end{proof}
\end{lemma}
\begin{remark}
Note that in order to represent $q-2$ auxiliary rows with
components in the form $\frac{\alpha _{i}^{(f-r)t+q-r}}{(\alpha
_{i}^{t+1}+1)^{q-1}}$ we used $q-3$ new auxiliary rows with
components in the form $\frac{\alpha
_{i}^{(f-r-1)t+q-r-1}}{(\alpha _{i}^{t+1}+1)^{q-2}}$.
\end{remark}
Recurrent equations for auxiliary rows with a reduced
denominator power of the components and the corresponding rows
of submatrices $h_{q-1}$ ,$h_{q-2}$, \ldots, $h_{q-f}$ can be
extended up to the last step for which we have the following
relation:
\begin{center}
$ \frac{\alpha _{i}^{t+q-(f-1)}}{(\alpha
_{i}^{t+1}+1)^{q-(f-1)}}= \frac{\alpha _{i}^{q-f}}{(\alpha
_{i}^{t+1}+1)^{q-f}}-\frac{\alpha _{i}^{q-f}}{(\alpha
_{i}^{t+1}+1)^{q-(f-1)}}$,
\end{center}
where the rows with components $\frac{\alpha
_{i}^{q-f}}{(\alpha _{i}^{t+1}+1)^{q-f}}$ , $\frac{\alpha
_{i}^{q-f}}{(\alpha _{i}^{t+1}+1)^{q-(f-1)}}$ belong to the
submatrices $h_{q-f}$ ,$h_{q-(f-1)}$.
Thus, with taking the above
statements into account, we can prove the following theorem.
\begin{theorem}
All rows of the submatrix $h_{1}$
\begin{equation*}
 \left[
 \begin{array}{ccccc}
 \frac{\alpha _{1}^{f\frac{t}{q}+1}}{\alpha _{1}^{t+1}+1} &
  \frac{\alpha _{2}^{f\frac{t}{q}+1}}{\alpha _{2}^{t+1}+1}&
 \ldots & \frac{\alpha _{n-1}^{f\frac{t}{q}+1}}{\alpha _{n-1}^{t+1}+1}& 0 \\
\end{array}
\right],
\end{equation*}
where $f=1,\ldots,q-1$, can be obtained    as a linear
combinations of the following rows of submatrices $h_{1},
h_{2}, \ldots, h_{q-1}$:
\begin{equation*}
\begin{array}{c}
 \left[
 \begin{array}{ccccc}
 \frac{\alpha _{1}^{t-f\frac{t}{q}}}{\alpha _{1}^{t+1}+1} &
  \frac{\alpha _{2}^{t-f\frac{t}{q}}}{\alpha _{2}^{t+1}+1}&
 \ldots & \frac{\alpha _{n-1}^{t-f\frac{t}{q}}}{\alpha _{n-1}^{t+1}+1}& 0 \\
\end{array}
\right] ,
\\
 \left[
 \begin{array}{ccccc}
 \frac{\alpha _{1}^{q-f}}{(\alpha _{1}^{t+1}+1)^{2}} &
  \frac{\alpha _{2}^{q-f}}{(\alpha _{2}^{t+1}+1)^{2}}&
 \ldots & \frac{\alpha _{n-1}^{q-f}}{(\alpha _{n-1}^{t+1}+1)^{2}}& 0 \\
\end{array}
\right] , \\
\vdots
\\
 \left[
 \begin{array}{ccccc}
 \frac{\alpha _{1}^{q-f}}{(\alpha _{1}^{t+1}+1)^{q-1}} &
 \frac{\alpha _{2}^{q-f}}{(\alpha _{2}^{t+1}+1)^{q-1}}&
 \ldots & \frac{\alpha _{n-1}^{q-f}}{(\alpha _{n-1}^{t+1}+1)^{q-1}}& 0 \\
\end{array}
\right] .
\end{array}
 \end{equation*}
\end{theorem}
It means that the number of $q$-ary rows equal to
$\theta_{q-1}=(q-1)2l$ can be additionally deleted from the
submatrix $h_{1}$. Thus, we get the following estimation for the
dimension of the cumulative-separable $(L, G)$-code with
$G_{6}^{(q)}(x) =\left(x^{t+1}+1\right)^{q}$ :
\begin{center}
$k_{6}^{(q)}\geq
n-2l\left((q-1)(t+1)-\frac{(q-1)(q+3)}{2}\right)$
\end{center}
or
\begin{equation}
k_{6}^{(q)}\geq
t^{2}-t-1-2l(q-1)\left(t-\frac{3}{2}-\frac{q-2}{2}\right).
\label{dimension_comulative_code_G_6^q}
\end{equation}

It should be noted that the previously constructed boundaries
for the dimensions of binary and $q$-ary separable codes of
this subclass are obtained from the above boundary as a special
case.
\begin{table}[H!]
	\caption{ Parameters of the cumulative-separable codes
$\Gamma_{6}^{(q)}=(x^{t+1}+1)^{q}$, $t=q^{l}$}

	\centering
		\begin{tabular}{|c|c|c|c|c|c|} \hline $q$ & $l$ & $n$ &
$\begin{array}{c}
   k\\
   \text{estimation} \\
   \eqref{dimension_comulative_code_G_6^q}
 \end{array}$
 & $\begin{array}{c}
   k\\
   \text{real value}
 \end{array}$
 &
  $\begin{array}{c}
   d \geq q(t+1)+1\\
   \text{estimation}
 \end{array}$
 \\ \hline
 3 & 2 & 71 & 15 & 16 & 31 \\ \hline
 3 & 3 &701 & 401& 401 & 85 \\ \hline
 3 & 4 & 6479 & 5215 & 5215 & 247 \\ \hline
 5 & 2 & 599 & 247 & 256 & 131 \\ \hline
 5 & 3 & 15499 & 12571 & 12571 & 631 \\ \hline
 7 & 2 & 2351 & 1271 & 1296 & 351  \\ \hline
 11 & 2 & 14519 & 9919 & 10000 & 1343 \\ \hline
		\end{tabular}
	\label{tab:Table_of_Comulative_Separable_codes_G6^q}
\end{table}
By using the technique of embedded code constructing described
in \cite{Bezz88} for the ternary cumulative-separable
$\Gamma_{6}(L_{6},G_{6}^{(3)})$-code  with
$L=\left\{GF(3^{4})\{\alpha: G_{6}(\alpha)=0\}\right\}$ and the
Goppa polynomial $G_{6}^{(3)}(x)=(x^{10}-1)^{3}$     we can
obtain the following collection of codes(Table
\ref{tab:Table_of_Embedded_comulative_separable_codes_G6^q}):
$$
\Gamma_{6}(L_{6},G_{6}^{(3)i}),
G_{6}^{(3)i}(x)=(x^{10}-1)^{3}(x-1)^{i}, \; i\geq 1 .
$$
Moreover,
the best codes from the known ternary block linear ones
\cite{Br} are present in this collection.
\begin{table*}[H!]
	\centering
	\caption{ Sequence of embedded ternary codes derived from
the code             $ \Gamma_{6}(L_{6},G_{6}^{(3)}),
G_{6}^{(3)}=(x^{10}-1)^{3}$ }

		\begin{tabular}{|c|c|c|c|c|c|} \hline
 $i$ & $G_{6}^{(3)i}=G_{6}^{(3)}(x-1)^{i}$ & $n$ & $k$ & $d$ & $d$ \cite{Br} \\ \hline
 0 & $(x^{10}-1)^{3}$ & 71 & 16 & 31 & 31 \\ \hline
 1 & $(x^{10}-1)^{3}(x-1)$ &\textbf{71} & \textbf{15} & \textbf{33} & \textbf{32} \\ \hline
 4 & $(x^{10}-1)^{3}(x-1)^{4}$ & 71 & 11& 35 & 36 \\ \hline
 10 & $(x^{10}-1)^{3}(x-1)^{10}$ & 71 & 7 & 42 & 42 \\ \hline
 13 & $(x^{10}-1)^{3}(x-1)^{13}$ & 71 & 5 & 44 & 45\\ \hline
		\end{tabular} 	
\label{tab:Table_of_Embedded_comulative_separable_codes_G6^q}
\end{table*}

\begin{table}[H!]
	\caption{ Parameters of the cumulative-separable codes
$\Gamma_{1}^{(q)}=(x^{t-1}+1)^{q}$, $t=q^{l}$}

	\centering
		\begin{tabular}{|c|c|c|c|c|c|} \hline $q$ & $l$ & $n$ &
$\begin{array}{c}
   k\\
   \text{estimation}\\
   (  \text{Section  \ref{subsection_chain_comulative_separablecodes}})
 \end{array}$
 & $\begin{array}{c}
   k\\
   \text{real value}
 \end{array}$
 &
  $\begin{array}{c}
   d \geq q(t-1)+1\\
   \text{estimation}
 \end{array}$
 \\ \hline
 3 & 2 & 73 & 16 & 17 & 25 \\ \hline
 3 & 3 &703 & 402& 402 & 79 \\ \hline
 3 & 4 & 6481 & 5216 & 5216 & 241 \\ \hline
 5 & 2 & 601 & 248 & 257 & 121 \\ \hline
 5 & 3 & 15501 & 12572 & 12572 & 621 \\ \hline
 7 & 2 & 2353 & 1272 & 1297 & 337  \\ \hline
 11 & 2 & 14522 & 9921 & 10002 & 1331 \\ \hline
		\end{tabular}
	\label{tab:Table_of_Comulative_Separable_codes_G1^q}
\end{table}
\begin{table}[h]
\caption{Dependence of cumulative-separable codes dimensions on
the cumulativity orders $1\leq i \leq q-1 $ for $q=3$, $l=2$
and $t= 9$ }

\begin{tabular}{|c|c|c|c|c|c|}
\hline  $i$& $
   \Gamma_{1}^{(i)}
 $
&
 $   \Gamma_{1}^{*(i)} $
& $   \Gamma^{(i)}_{2} \equiv \Gamma^{(i)}_{3}$
& $
   C^{*(i)}_{3} \equiv \Gamma^{*(i)}_{4} $
&
 $  \Gamma^{(i)}_{5} \equiv \Gamma^{(i)}_{6} $
 \\

   &$n_{1}=73$  & $n_{1}^{*}=72$ & $n_{2}=72$ & $n_{3}^{*}=71$  &
   $n_{5}=71$
   \\ \hline
 1 & 42 & 41 & 39 & 39 &  37  \\ \hline
 2 & 17 & 16 & 16 & 16 &  16  \\ \hline
\end{tabular}
\end{table}

\begin{table}[h]
\caption{ Dependence of  cumulative-separable codes dimensions
on the
 cumulativity orders $1\leq i \leq q-1$ for
$q=3$, $l=3$ and $t= 27$}
\begin{tabular}{|c|c|c|c|c|c|}

\hline  $i$& $
   \Gamma_{1}^{(i)}
 $
&
 $   \Gamma_{1}^{*(i)} $
& $   \Gamma^{(i)}_{2} \equiv \Gamma^{(i)}_{3}$ & $
   C^{*(i)}_{3} \equiv \Gamma^{*(i)}_{4} $
&
 $  \Gamma^{(i)}_{5} \equiv \Gamma^{(i)}_{6} $
 \\

   &$n_{1}=703$  & $n_{1}^{*}=702$ & $n_{2}=702$ & $n_{3}^{*}=701$  &
   $n_{5}=701$
   \\ \hline
 1 & 549 & 548 & 545 & 545 &  542  \\ \hline
 2 & 402 & 401 & 401 & 401 &  401  \\ \hline
\end{tabular}
\end{table}

\begin{table}[h]
\caption{ Dependence of  cumulative-separable codes dimensions
on the cumulativity orders $1\leq i \leq q-1$ for $q=5$, $l=2$
and $t= 25$}
\begin{tabular}{|c|c|c|c|c|c|}

\hline  $i$& $
   \Gamma_{1}^{(i)}
 $
&
 $   \Gamma_{1}^{*(i)} $
& $   \Gamma^{(i)}_{2} \equiv \Gamma^{(i)}_{3}$ & $
   C^{*(i)}_{3} \equiv \Gamma^{*(i)}_{4} $
&
 $  \Gamma^{(i)}_{5} \equiv \Gamma^{(i)}_{6} $
 \\

   &$n_{1}=601$  & $n_{1}^{*}=600$ & $n_{2}=600$ & $n_{3}^{*}=599$  &
   $n_{5}=599$
   \\ \hline
 1 & 506 & 505 & 503 & 503 &  501  \\ \hline
 2 & 412 & 411 & 409 & 409 &  407  \\ \hline
 3 & 322 & 321 & 319 & 319 &  317  \\ \hline
 4 & 257 & 256 & 256 & 256 &  256  \\ \hline
\end{tabular}
\end{table}

\begin{table}[h]
\caption{ Dependence of cumulative-separable codes dimensions
on the
 cumulativity orders $1\leq i \leq q-1$ for
$q=7$, $l=2$ and $t= 49$}
\begin{tabular}{|c|c|c|c|c|c|}

\hline  $i$& $
   \Gamma_{1}^{(i)}
 $
&
 $   \Gamma_{1}^{*(i)} $
& $   \Gamma^{(i)}_{2} \equiv \Gamma^{(i)}_{3}$ & $
   C^{*(i)}_{3} \equiv \Gamma^{*(i)}_{4} $
&
 $  \Gamma^{(i)}_{5} \equiv \Gamma^{(i)}_{6} $
 \\
 &$n_{1}=2353$  & $n_{1}^{*}=2352$ & $n_{2}=2352$ & $n_{3}^{*}=2351$  &
   $n_{5}=2351$
   \\ \hline

 1 & 2162 & 2161 & 2159 & 2159 &  2157  \\ \hline
 2 & 1972 & 1971 & 1969 & 1969 &  1967  \\ \hline
 3 & 1786 & 1785 & 1783 & 1783 &  1781  \\ \hline
 4 & 1604 & 1603 & 1601 & 1601 &  1599  \\ \hline
 5 & 1426 & 1425 & 1423 & 1423 &  1421  \\ \hline
 6 & 1297 & 1296 & 1296 & 1296 &  1296  \\ \hline
\end{tabular}
\end{table}

\end{document}